\documentclass[pra,twocolumn,amsmath,amssymb,aps,preprintnumbers,superscriptaddress]{revtex4-1}

\usepackage{amsmath, amsfonts, amssymb}
\usepackage{graphicx}
\usepackage{hyperref}
\usepackage[english]{babel}

\begin{document}

\title{Constraints on short-range gravity with self-gravitating Bose-Einstein condensates}
\author{S. G\"odtel}
\author{C. L\"ammerzahl} 
\affiliation{ZARM, University of Bremen, Germany}
\date{April 3, 2023}

\begin{abstract}
In this work, we study low-lying collective excitations of a Bose-Einstein condensate with Newtonian and Yukawa-like two-particle interaction and derive boundaries for both Yukawa parameters. Using a variational approach, we explicitly show for spherical condensate that the corresponding frequencies depend on the gravitational interaction strength. The acquired results are presented in contour plots and compared to experimentally verified data from other tests. Furthermore, we discuss experimental requirements to test our theoretical model as well as possibilities to improve the boundaries. In addition, we consider axisymmetric condensates, where it turns out that disk-shaped BECs lead to better constraints. We also show that in theory we can determine the values for both Yukawa parameters independently by a measurement of at least two collective frequencies. 
\end{abstract}

\maketitle

\section{Introduction\label{sec:intro}}

\begin{figure}[t!]
	\includegraphics[scale=0.45]{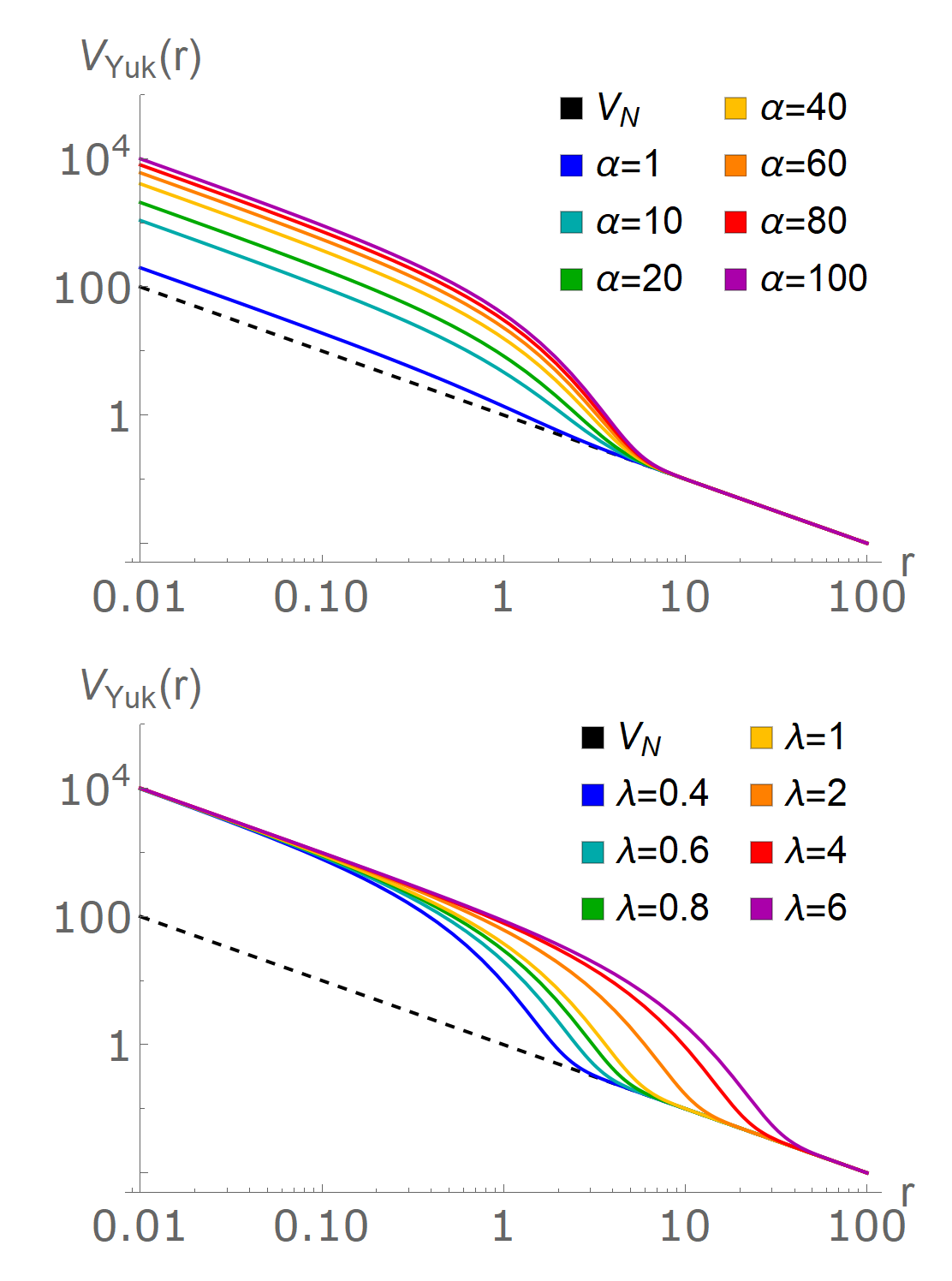}
	\caption{(Colors online) Schematic sketch of the Yukawa potential. On the upper picture we choose $\lambda=1$ and in the lower figure $\alpha=100$. The dashed black line denotes the Newtonian potential.}
	\label{fig:Yuk parameters}
\end{figure}

General relativity and the Standard Model are two of the best descriptions of nature available to us to date, and both of them have been consistently verified in the experiment. On one side, the exact perihelion shift of Mercury~\cite{Will} and the existence of gravitational waves~\cite{LIGO} can be explained, while on the other side top quarks~\cite{Abachi} and Higgs bosons~\cite{ATLAS} have been detected. However, both theories are incompatible due to the lack of a quantum description of gravity and the hierarchy problem, the relative weakness of gravity compared to other fundamental forces. As a consequence, many theories have been proposed in recent decades addressing modifications to Newton's law of gravity, in particular at short-ranges. The explanation of such modifications range from a finite interaction with an additional dilaton field~\cite{Fuji}, a distance-dependent gravitational constant~\cite{Long}, and a fifth force\cite{Fischbach} to the predictions of extra dimensions~\cite{Arkani, Arkani2}. In many cases the Newtonian gravitational potential
\begin{align}\label{eq:Newton pot}
V_\mathrm{N}(r) = -G\frac{Mm}{r}
\end{align}
is modified in the submillimeter regime and then parametrized in the form of a Yukawa potential
\begin{align}\label{eq:Yuk pot}
V_\mathrm{Yuk} = -G\frac{Mm}{r}\left(1+\alpha\exp\left\{-\frac{r}{\lambda}\right\}\right).
\end{align}
Here $G$ denotes the gravitational constant, $M$ and $m$ the masses of the two interacting bodies and $r$ the distance between the two bodies. Furthermore, the Yukawa potential introduces two additional degrees of freedom: the interaction strength $\alpha$ and the effective range $\lambda$. The effects of both parameters are shown in Fig.~\ref{fig:Yuk parameters}. From a physical point of view, the effective range could be the Compton wavelength of an exotic particle or the radius of the compactification of extra dimensions.

The amount of theoretical predictions led to numerous experimental tests of Newton's law of universal gravitation. These include collider experiments with proton-proton collisions~\cite{Aaltonen, Aad, Aad2}, Casimir forces~\cite{Lamoreaux, Lamoreaux2, Mohideen}, cantilever tests~\cite{Smullin, Chiaverini}, torsion balance pendulum~\cite{Adelberger2, Kapner} to astronomical observations~\cite{Iorio}. In total, this covers an astonishing range over more than 35 orders of magnitude for both the effective range and the interaction strength. For a comprehensive collection, we recommend Refs.~\cite{Murata, Adelberger}. Surprisingly, no deviations from Newtons inverse square law have been found so far. As a result, one usually finds exclusion diagrams including constraints for the Yukawa parameters. The search for Non-Newtonian physics has therefore become the task of improving the constraints by upgrading existing experiments and develop new tests.

For the purpose of this paper, we introduce here a theoretical concept of a self-gravitating Bose-Einstein condensate. This was first discussed by Ruffini and Bonazola~\cite{Ruffini} in an astrophysical context and proposed as a hypothetical object known as a boson star~\cite{Feinblum, Colpi}. In some proposals, these condensates also serve as dark matter candidates~\cite{Harko}. From a theoretical point of view, bosons are described by a complex scalar field satisfying the generalized Einstein-Klein-Gordon system. This system has been extensively studied, see e.g., Ref.~\cite{Jetzer}. A nonrelativistic limit is known as the Gross-Pitaevksii-Newton system given by
\begin{align}\label{eq:GPNE}
i\hbar\;\partial_t\Psi &= -\frac{\hbar^2}{2 m}\Delta\Psi + V\Psi + g\vert\Psi\vert^2\Psi + m\Phi\Psi, \nonumber\\
\Delta\Phi &= 4\pi G\vert\Psi\vert^2.
\end{align}
Here $V$ denotes an external potential, $g\vert\Psi\vert^2$ the usual two-particle contact interaction, and $m\Phi$ a gravitational potential satisfying the Poisson equation. Among others, the collapse~\cite{Chavanis}, stable solutions~\cite{Schroven, Chavanis2} and the Thomas-Fermi limit~\cite{Toth} were discussed. In addition, numerical programs have been developed based on the Crank-Nicholson method~\cite{Madarassy} and the Gross-Pitaevskii-Newton system was also applied to ultracold plasma~\cite{Sakaguchi} and dipolar Bose-Einstein condensates~\cite{Bao}.

In this work, we propose a theoretical model of a self-gravitating Bose-Einstein condensate as an additional test of modified gravity. In such a condensate, we include a gravitational two-particle interaction in addition to the commonly used contact interaction. As two examples, we consider a Newtonian and a Yukawa-like potential. In the case of the latter, we look for constraints for both parameters, the strength and the effective range. We explicitly aim for a theoretical improvement of the constraints measured with Casimir forces~\cite{Ederth, Harris} and electron spin precession~\cite{Tanaka}, i.e. for an effective range in the submillimeter regime and an interaction strength smaller than $10^{30}$. In contrast to most experiments, which focus on the influence of external gravitational fields, we study a quantum many-particle system with an intrinsic gravitational interaction. To encourage experimental verification, we study low-lying collective frequencies of such a condensate, since these can be measured with a relative precision of $10^{-3}$~\cite{Stamper, You}.

This work is organized as follows. As the theoretical foundation, we present in Sec.~\ref{sec:var} a variational approach introduced in Ref.~\cite{Perez}. Although we strictly follow the steps explained there, we generalize here the expressions to arbitrary two-particle interactions and any symmetry. By choosing an ansatz for the condensate wave function in form of a Gaussian function, we derive an expression for the differential equations describing time-dependent changes of the Gaussian width. This allows us to find a steady-state as well as Hessian matrix via a small perturbation out of the equilibrium. The eigenvectors of this matrix represent the collective modes and the eigenvalues lead to the corresponding collective frequencies. 
In the following Sec.~\ref{sec:general}, we specify the type of the two-particle interaction. We start with the local contact interaction, followed by two models of long-range gravitational interaction, namely a Newtonian and a Yukawa-like interaction, given by the Eqs.~\eqref{eq:Newton pot} and~\eqref{eq:Yuk pot}, respectively. Based on the general equations derived in the previous section, we determine symmetry independent formulas for all three interactions. Since the gravitational interactions diverge at the origin, we decide to apply a Fourier transformation analoguous to recent theoretical discussions concerning dipolar condensates, see e.g., Ref.~\cite{Muruganandam}.
Next, we specify the symmetry of the condensate and present in Sec.~\ref{sec:sphere} the results for a spherically symmetric condensate as the simplest case, followed by a generalization to axially symmetric condensates in Sec.~\ref{sec:axial}. While in case of the contact interaction our results coincide with the literature~\cite{Perez}, we report analytical expressions for the equilibrium cloud width and the Hessian matrix for both Newtonian and Yukawa-like interactions. It turns out that in case of Newtonian interaction, the corrections to the collective frequencies are insignificant for real applications in the laboratory. However, if we assume Yukawa-like interactions, we obtain expressions containing two additional parameters: the interaction strength and the effective range. Accordingly, our results are presented as contour plots comparable to experimentally verified data of different setups. Moreover, we show the influence of accessible experimental parameters to find the best possible constraints.
Finally, we summarize our results in Sec.~\ref{sec:summary} and give a short outlook.

\section{Variational method}\label{sec:var}

We consider a Bose-Einstein condensate at zero temperature confined in a harmonic trap potential
\begin{align}\label{eq:3D harmonic trap}
V(\textbf{r}) = \frac{m}{2}\omega^2(\nu_x^2 x^2+\nu_y^2 y^2+\nu_z^2 z^2).
\end{align}
Here $\omega$ serves as a frequency scale and with the dimensionless numbers $\nu_j$ we specify later on the symmetry of the trap and thus the symmetry of the condensate.

As stated in Ref.~\cite{Perez}, the solution of the time-dependent Gross-Pitaevskii equation can also be formulated as a variational problem. There we have to minimize the corresponding Lagrangian density
\begin{align}\label{eq:Lagrange density}
\mathcal{L} = i\hbar\Psi^\star\partial_t\Psi - \frac{\hbar^2}{2m}\nabla\Psi^\star\nabla\Psi - V\Psi^\star\Psi - V_\mathrm{int}\Psi^{\star 2}\Psi^2
\end{align}
according to Hamilton's principle. Here $V_\mathrm{int}$ denotes an arbitrary two-particle interaction potential. To obtain a dynamic result for the condensate wave function $\Psi$ and its complex conjugate $\Psi^\star$, we need to choose a suitable test function. Since we are interested in low-lying oscillations around the ground state, a natural choice is a generalized Gau\ss ansatz in the form
\begin{align}\label{eq:Gauss ansatz}
\Psi(\textbf{x},t) = &\frac{\sqrt{N}}{\sqrt[4]{\pi^3}\sqrt{A_1(t) A_2(t) A_3(t)}} \nonumber\\
&\times\exp\left\{-\sum_{j=1}^3\left(\frac{1}{2A_j(t)^2}+iB_j(t)\right)x_j^2\right\}
\end{align}
since we know that in the limiting case without particle interaction, the Gross-Pitaevskii equations reduces to a linear Schr\"odinger equation, which itself is solved by a Gaussian. The ansatz~\eqref{eq:Gauss ansatz} is chosen in such a way that the wave function is normalized to the particle number such that
\begin{align}\label{eq:norm wave function}
\int\mathrm{d}x^3\vert\Psi\vert^2 = N.
\end{align}
Furthermore, the parameters $A_j(t)$ describe the widths and $B_j(t)$ the expansion or contraction velocities, respectively, of the Gauss function in each spatial direction. Note that the inclusion of the parameters $B_j(t)$ is essential to get the correct dynamical behavior. In the following, both sets $A_j(t)$ and $B_j(t)$ will be used as the variational parameters. For better readability, we omit the time-dependency in the notation from here on.

If we now insert the Gau\ss ansatz~\eqref{eq:Gauss ansatz} into the Lagrange density~\eqref{eq:Lagrange density}, we obtain a Lagrangian depending on the variational parameters $A_j$ and $B_j$ by integrating over the spatial coordinates
\begin{align}\label{eq:L complete}
L= \sum_{j=1}^3\Bigg[ &\frac{\hbar}{2}N A_j^2\dot{B}_j -\frac{\hbar^2}{m}N \left(\frac{1}{4A_j^2}+A_j^2B_j^2\right) \nonumber\\
& -\frac{m}{4}N \omega^2\nu_j^2A_j^2\Bigg] + L_\mathrm{int}.
\end{align}
We retain here the interaction term $L_\mathrm{int}$ in a general form defined by
\begin{align}\label{eq:Lint def}
L_\mathrm{int} = -\int\mathrm{d}^3x\int\mathrm{d}^3x'\;\vert\Psi(\textbf{x},t)\vert^2 V_\mathrm{int}(\textbf{x}-\textbf{x}')\vert\Psi(\textbf{x}',t)\vert^2.
\end{align}
Specifications of the interaction potential $V_\mathrm{int}$ are discussed in later sections. However, using the Gau\ss ansatz, it turns out that $L_\mathrm{int}$ is in general independent of the parameters $B_j$, since
\begin{align}
L_\mathrm{int} = &-\frac{N^2}{\pi^3} \frac{1}{A_1^2A_2^2A_3^2} \nonumber\\
&\times\int\mathrm{d}^3x\int\mathrm{d}^3x'\;V_\mathrm{int}(\textbf{x}-\textbf{x}')\exp\left\{-\sum_{j=1}^3\frac{x_j^2-x_j'^2}{A_j^2}\right\}.
\end{align}

Now we minimize the Lagrangian in Eq.~\eqref{eq:L complete} with respect to the variational parameters $A_j$ and $B_j$. The Euler-Lagrange equations for $B_j$ can be inserted into the equations for $A_j$, yielding a set of three differential equations of second order
\begin{align}
\ddot{A}_j+\omega^2\nu_j^2A_j = \frac{\hbar^2}{m^2}\frac{1}{A_j^3} + \frac{2}{mN}\;\partial_{A_j}L_\mathrm{int}
\end{align}
describing the evolution of the widths of the condensate in each spatial direction. To put these expressions in a more compact form, we introduce the dimensionless units $\tau=\omega t$ and $\gamma_j = A_j/l$, where $l=\sqrt{\hbar/(m\omega)}$ denotes the oscillator length. Thus the differential equations read
\begin{align}\label{eq:ODE general}
\ddot{\gamma}_j = &-\nu_j^2\gamma_j + \frac{1}{\gamma_j^3} + \frac{2}{N\hbar\omega}\;\partial_{\gamma_j} L_\mathrm{int}.
\end{align}
Note that the interaction term of the Lagrangian now depends on the dimensionless Gauss widths $\gamma_j$
\begin{align}
L_\mathrm{int} = L_\mathrm{int}\left(l \gamma_x, l \gamma_y, l \gamma_z\right).
\end{align}

As discussed in Ref.~\cite{Perez}, the differential equations in~\eqref{eq:ODE general} resemble harmonic oscillators with a dispersive kinetic term and a nonlinear interaction term. Consequently, we interpret the equations as the classical motion of a point particle in an effective potential given by the negative derivative of the right-hand side of Eq.~\eqref{eq:ODE general}. Thus we define
\begin{align}\label{eq:eff pot general}
V_\mathrm{eff}(\gamma_x,\gamma_y,\gamma_z) = \frac{1}{2}\sum_j\left(\nu_j^2\gamma_j^2 +\frac{1}{\gamma_j^2} \right) - \frac{2}{N\hbar\omega}L_\mathrm{int}.
\end{align}

The low-lying excitations are small collective oscillations of the condensate around an equilibrium width. This equilibrium width is found either as the minimum of the effective potential~\eqref{eq:eff pot general} or by setting the acceleration term in the differential equation~\eqref{eq:ODE general} equal to zero and solving the algebraic equations. Both calculations lead to the same result for the equilibrium width, as we will explicitly show in later sections. Assuming a small perturbation out of the equilibrium, we apply a Taylor expansion up to second order to the effective potential. The first order term cancels out due to the condition of the equilibrium and the second order term is proportional to the Hessian matrix
\begin{align}\label{eq:M def}
M = \Big( \left.\partial_{\gamma_j}\partial_{\gamma_k} V_\mathrm{eff}\right\vert_{\gamma=\gamma_0} \Big)_{j,k},
\end{align}
which contains the second derivatives of the effective potential. In general, the eigenvectors are the collective modes and the square root of the eigenvalues leads to the ratio of the collective frequency and the frequency scale $\Omega/\omega$.

For simplicity, we now decompose the Hessian matrix into the sum
\begin{align}\label{eq:M general}
M = M_1 + M_\mathrm{int},
\end{align}
which contains a single-particle contribution
\begin{align}\label{eq:M1 general}
M_1 =
\begin{pmatrix}
\nu_x^2+\frac{3}{\gamma_{x0}^4} & 0 & 0 \\
0 & \nu_y^2+\frac{3}{\gamma_{y0}^4} & 0 \\
0 & 0 & \nu_z^2+\frac{3}{\gamma_{z0}^4} 
\end{pmatrix}
\end{align}
including the kinetic and the trapping part, and a contribution due to the two-particle interaction
\begin{align}\label{eq:Mint general}
M_\mathrm{int} = -\frac{2}{N\hbar\omega}\;\Bigg( \left.\partial_{\gamma_j}\partial_{\gamma_k} L_\mathrm{int}\right\vert_{\boldsymbol{\gamma} = \boldsymbol{\gamma}_0} \Bigg)_{j,k}.
\end{align}
Note that we first have to perform the derivatives of the Lagrangian and after that we evaluate the result at the equilibrium.

\section{General considerations}\label{sec:general}

In this section, we specify the type of the interaction and derive symmetry independent equations.

\subsection{Contact interaction}\label{sec:general con}

As the first example of a two-particle interaction, we assume the commonly used contact interaction potential
\begin{align}\label{eq:con pot}
V_\mathrm{con}(\textbf{x}-\textbf{x}') = g\delta(\textbf{x}-\textbf{x}'),
\end{align}
where the self-interaction factor $g=4\pi\hbar^2a_\mathrm{s}/m$ contains the s-wave scattering length $a_\mathrm{s}$ and the mass $m$ of the particles forming the condensate. Interestingly, the sign of the scattering length can be addressed experimentally via Feshbach resonances~\cite{Pitaevskii2, Chin} and determines whether the interaction is attractive ($a_\mathrm{s}<0$) or repulsive ($a_\mathrm{s}>0$). Further details can be found in Ref.~\cite{Perez2}. However, in this work we only consider positive scattering lengths and thus repulsive contact interactions.

With the contact potential~\eqref{eq:con pot} it is easy to derive the corresponding Lagrangian. Applying the Gaussian ansatz in Eq.~\eqref{eq:Gauss ansatz} leads to
\begin{align}\label{eq:Lcon}
L_\mathrm{con} = -\frac{gN^2}{2\sqrt{(2\pi)^3}}\left(\sqrt{\frac{m\omega}{\hbar}}\right)^3 \frac{1}{\gamma_x\gamma_y\gamma_z},
\end{align}
expressed in dimensionless units. This immediately results in the effective potential
\begin{align}\label{eq:Veff con general}
V_\mathrm{eff}^{\mathrm{(con)}} = \frac{1}{2}\sum_j\left(\nu_j^2\gamma_j^2+\frac{1}{\gamma_j^2}\right) + \frac{P}{\gamma_x\gamma_y\gamma_z},
\end{align}
where we define the dimensionless contact interaction strength
\begin{align}\label{eq:def P}
P = \frac{gN}{\sqrt{(2\pi)^3}}\frac{m}{\hbar^2}\sqrt{\frac{m\omega}{\hbar}} = \sqrt{\frac{2}{\pi}}N \frac{a_\mathrm{s}}{l}.
\end{align}

According to Eq.~\eqref{eq:ODE general} we also derive the differential equations
\begin{align}\label{eq:ODE con}
\ddot{\gamma}_j = -\nu_j^2\gamma_j + \frac{1}{\gamma_j^3} + \frac{P}{\gamma_x\gamma_y\gamma_z}\frac{1}{\gamma_j}.
\end{align}
As mentioned in the previous section, the equilibrium cloud widths are determined by setting the acceleration in the differential equation~\eqref{eq:ODE con} equal to zero, so that we have
\begin{align}\label{eq:equi width con}
-\nu_j^2\gamma_{j0} + \frac{1}{\gamma_{j0}^3} + \frac{P}{\gamma_{x0}\gamma_{y0}\gamma_{z0}}\frac{1}{\gamma_{j0}} = 0.
\end{align}

Furthermore, the contribution to the diagonal and off-diagonal elements of the Hessian matrix due to the contact interaction is given by
\begin{align}\label{eq:Mcon}
M_\mathrm{con}^{(jj)} &= \frac{P}{\gamma_{x0}\gamma_{y0}\gamma_{z0}} \frac{2}{\gamma_{j0}^2}, \\
M_\mathrm{con}^{(jk)} &= \frac{P}{\gamma_{x0}\gamma_{y0}\gamma_{z0}} \frac{1}{\gamma_{j0}\gamma_{k0}}
\end{align}
with $j\neq k$. The expressions so far are valid in any symmetry because of the local nature of the contact interaction.

\subsection{Newtonian interaction}\label{sec:general Newton}

Now in order to include a gravitational effects caused by the gravitational interaction between the particles, we now consider, in addition to the contact interaction, a Newtonian two-particle interaction given by~\eqref{eq:Newton pot}. Note that for a condensate with one atomic species, the masses are equal. Since the potential itself has a singularity at the origin, we choose to derive the Lagrangian in Fourier space. This method has already been successfully applied in the context of dipolar condensates for an interaction proportional to $r^{-3}$, see Ref.~\cite{Muruganandam}. The Lagrangian of the Newtonian interaction is then
\begin{align}\label{eq:LN FT}
L_\mathrm{N} = -\frac{1}{2}\frac{1}{(2\pi)^3}\int\mathrm{d}^3k\;\tilde{n}(\textbf{k})\;\tilde{V}_\mathrm{N}(\textbf{k})\;\tilde{n}(-\textbf{k})
\end{align}
with the Fourier transformed density
\begin{align}\label{eq:density FT}
\tilde{n}(\textbf{k}) = N\exp\left\{-\frac{1}{4}(A_x^2 k_x^2+A_y^2 k_y^2+A_z^2 k_z^2)\right\}
\end{align}
according to the Gaussian ansatz~\eqref{eq:Gauss ansatz} and the Fourier transformed Newtonian potential
\begin{align}\label{eq:Newton pot FT}
\tilde{V}_\mathrm{N}(\textbf{k})= -u\frac{4\pi}{k^2}
\end{align}
with the short-hand notation $u=Gm^2$. Using dimensionless units we then obtain general
\begin{align}\label{eq:LN dimless}
L_\mathrm{N} = &\frac{uN^2}{(2\pi)^2}\sqrt{\frac{m\omega}{\hbar}} \nonumber\\
&\times\int\mathrm{d}^3\kappa\;\frac{1}{\kappa^2}\exp\left\{-\frac{1}{2}(\gamma_x^2\kappa_x^2+\gamma_y^2\kappa_y^2+\gamma_z^2\kappa_z^2)\right\},
\end{align}
which is valid for any symmetry. However, in contrast to the local contact interaction, to our knowledge this integral cannot be solved in general. Therefore, we derive the differential equations and the Hessian matrix for a specific symmetry in the corresponding sections.

\subsection{Yukawa interaction}\label{sec:general Yuk}

Analogous to the Newtonian interaction, we apply the Fourier transformation to the Yukawa-like interaction potential given in~\eqref{eq:Yuk pot}. This leads to
\begin{align}\label{eq:Yukawa pot FT}
\tilde{V}_\mathrm{Yuk}(\textbf{k}) = -u\frac{4\pi}{k^2} - \alpha u\frac{4\pi}{k^2+\frac{1}{\lambda^2}},
\end{align}
which is then used to determine the Lagrangian of the Yukawa-like interaction
\begin{align}\label{eq:LYuk dimless}
L_\mathrm{Yuk} = \frac{uN^2}{(2\pi)^2} & \sqrt{\frac{m\omega}{\hbar}} \int\mathrm{d}^3\kappa\;\left( \frac{1}{\kappa^2}+\frac{\alpha}{\kappa^2+\frac{1}{\bar{\lambda}^2}} \right) \nonumber\\
&\times\exp\left\{-\frac{1}{2}(\gamma_x^2\kappa_x^2+\gamma_y^2\kappa_y^2+\gamma_z^2\kappa_z^2)\right\}
\end{align}
in dimensionless units. Note that $\bar{\lambda}=\lambda/l$ denotes the dimensionless effective range. To solve these integrals, we again need to specify the symmetry, which we discuss below.

\section{Spherical condensates}\label{sec:sphere}

As the theoretically simplest case, we consider in this section spherically symmetric condensates. These are realized by setting the trap frequencies $\nu_j$ equal in each spatial direction. As a result, the dimensionless Gaussian widths also coincide. Without loss of generality, we choose $\nu=1$ here. In the following, we discuss each interaction separately and show concrete results.

\subsection{Contact interaction}\label{sec:sphere con}

\begin{figure}[t!]
	\includegraphics[scale=0.45]{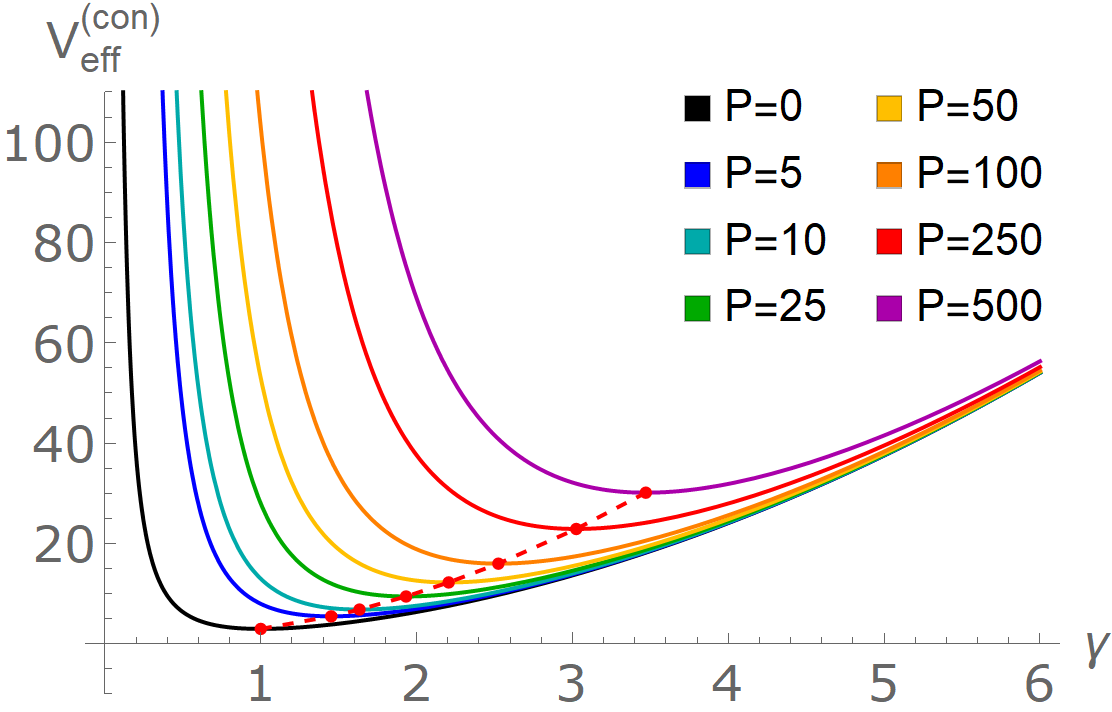}
	\caption{(Colors online) Effective potential $V_\mathrm{eff}^\mathrm{(con)}$ as a function of the Gaussian width $\gamma$ for a spherically symmetric condensate. The curves correspond to different contact interaction strength $P$ as indicated by the color. For each curve we include the minimum as well as a interpolation given by the red dashed line.}
	\label{fig:Veff con}
\end{figure}

For the contact interaction, we can simply read off the effective potential from Eq.~\eqref{eq:Veff con general}. In spherical symmetry this leads to
\begin{align}\label{eq:Veff con}
V_\mathrm{eff}^{\mathrm{(con)}} = \frac{3}{2}\left(\gamma^2+\frac{1}{\gamma^2}\right) + \frac{P}{\gamma^3}.
\end{align}
This function is shown in Fig.~\ref{fig:Veff con} depending on the Gaussian width for different values of the contact interaction strength. It is easy to find the minimum of each curve, indicated by red dots. This minimum corresponds to the equilibrium cloud width, and as can be seen in the figure, this width is $1$ for no particle interaction and increases with increased interaction strength, indicating a higher repulsion between the particles. A similar picture is also shown in Ref.~\cite{Perez}.

\begin{figure}[t!]
	\includegraphics[scale=0.45]{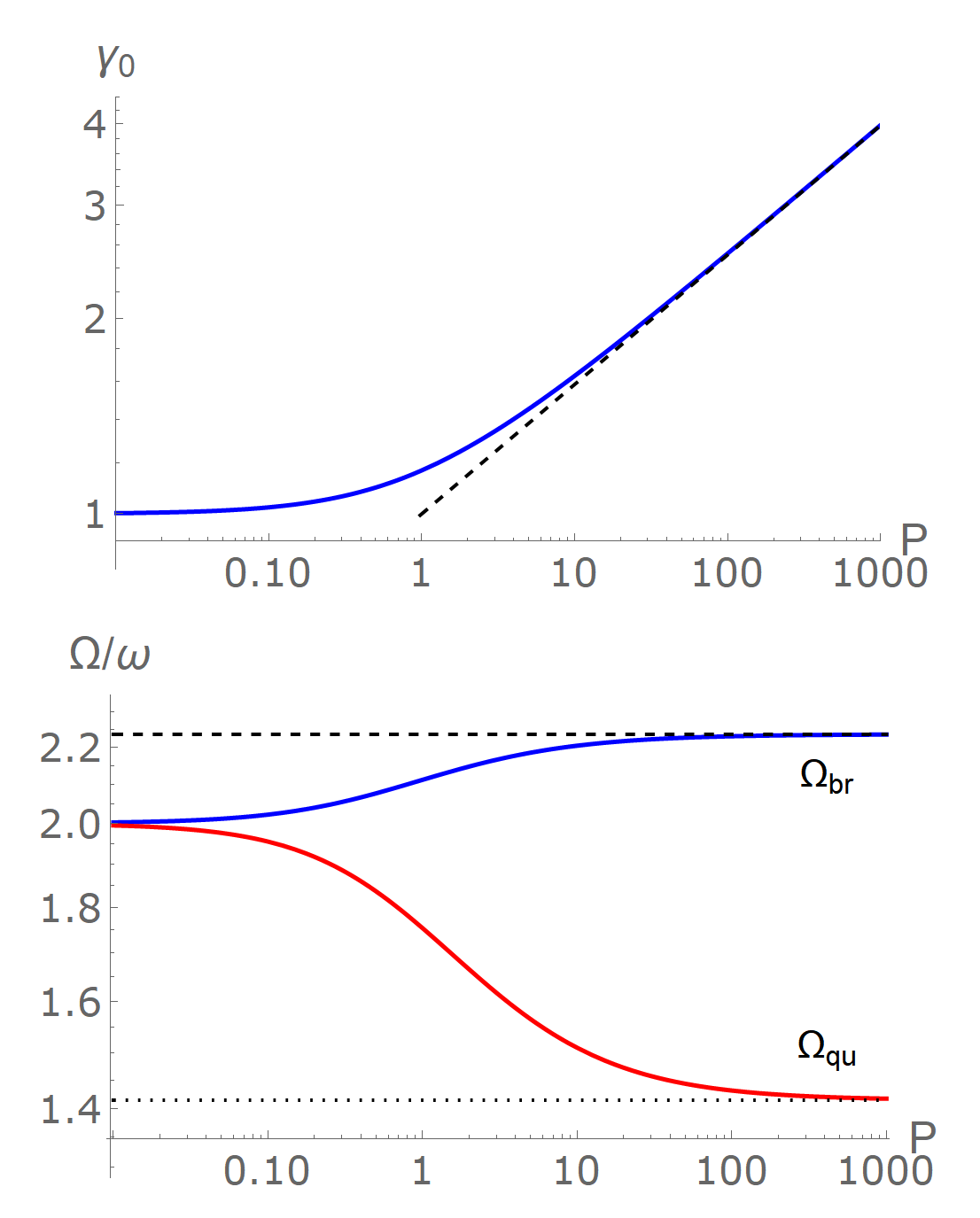}
	\caption{(Colors online) Equilibrium cloud width $\gamma_0$ and collective frequencies $\Omega_\mathrm{br}^\mathrm{(con)}/\omega$ and $\Omega_\mathrm{qu}^\mathrm{(con)}/\omega$ of a spherical condensate depending on the contact interaction strength $P$. The black dashed and dotted lines correspond to the Thomas-Fermi limits, respectively.}
	\label{fig:sphere con}
\end{figure}

As mentioned previously, we can also determine the equilibrium via the differential equations. In the special case of spherical symmetry, the three equations in~\eqref{eq:ODE con} reduce to only one. If we set the acceleration to zero, we get a single algebraic equation
\begin{align}\label{eq:equi width con sphere}
\gamma_0^5 - \gamma_0 - P = 0
\end{align}
for the equilibrium cloud width. This equation is solved numerically and the results are shown in Fig.~\ref{fig:sphere con}. The values agree with the minima of the effective potential, confirming that this is indeed an alternative way to calculate the equilibrium cloud width. 

With the equilibrium cloud width determined, we now apply the spherical symmetry to the Hessian matrix given by~\eqref{eq:M general} with the contribution of the contact interaction in Eq.~\eqref{eq:Mcon}. Its eigenvalues then lead to the collective frequencies
\begin{align}
\frac{\Omega_\mathrm{br}^\mathrm{(con)}}{\omega} = \sqrt{1+\frac{3}{\gamma_0^4}+\frac{4P}{\gamma_0^5}}, \label{eq:coll frequ con sphere 1}\\
\frac{\Omega_\mathrm{qu}^\mathrm{(con)}}{\omega} = \sqrt{1+\frac{3}{\gamma_0^4}+\frac{P}{\gamma_0^5}}. \label{eq:coll frequ con sphere 2}
\end{align}
The first expression is the frequency of a radial oscillation, which is therefore called the breathing mode. The second expression refers to a degenerate eigenvalue corresponding to the frequency of two quadrupole modes. Both frequencies are shown in the lower panel of Fig.~\ref{fig:sphere con} depending on the contact interaction strength. Without any particle interaction, the frequencies are degenerate at $\Omega/\omega=2$ and split up for nonzero interaction strengths. The frequency of the breathing mode is increased and that of the quadrupole modes is decreased. Both the equilibrium cloud width and the collective frequencies approach a certain asymptote for large interaction strengths. This asymptote is known as the Thomas-Fermi limit, see~\cite{Pethick}. For large two-particle interactions the kinetic contribution can be neglected, leading to the simple relation that $\gamma_0$ is equal to the fifth root of the interaction strength $P$. This relation is shown as a black dashed line in the upper picture. On the other hand, the Thomas-Fermi limits of the collective frequencies are constants given by $\Omega_\mathrm{br}^{\mathrm{(con)}}/\omega = \sqrt{5}$ and $\Omega_\mathrm{qu}^{\mathrm{(con)}}/\omega = \sqrt{2}$, respectively.

For later sections, we mention here a typical value for the contact interaction strength. Assuming a commonly used $^{87}\mathrm{Rb}$ condensate with $10^5$ particles, a s-wave scattering length of $90a_0$~\cite{Pethick}, where $a_0$ denotes the Bohr radius, and an angular trap frequency of $1\;\mathrm{kHz}$, the interaction strength is $P\approx 446$. Compared to Fig.~\ref{fig:sphere con}, this is clearly in the Thomas-Fermi regime.

\subsection{Newtonian interaction}\label{sec:sphere Newton}

Based on the symmetry independent expression for the Lagrangian of the Newtonian interaction~\eqref{eq:LN dimless}, we now consider spherical symmetry to explicitly derive corrections to the equilibrium cloud width and the collective frequencies due to the gravitational interaction.

We begin with the effective potential. For this purpose we evaluate the integrals in~\eqref{eq:LN dimless} in spherical coordinates. The calculation is straightforward and eventually leads to the effective potential
\begin{align}\label{eq:Veff Newton}
V_\mathrm{eff}^{\mathrm{(N)}} = \frac{3}{2}\left(\gamma^2+\frac{1}{\gamma^2}\right) + \frac{P}{\gamma^3} - \frac{3Q}{\gamma}
\end{align}
containing a contribution due to a Newtonian particle interaction. Here we define a dimensionless gravitational interaction strength
\begin{align}\label{eq:def Q sphere}
Q = \frac{1}{3}\sqrt{\frac{2}{\pi}}\frac{uN}{\hbar\omega}\sqrt{\frac{m\omega}{\hbar}} = \frac{1}{3}\sqrt{\frac{2}{\pi}} N\frac{a_\mathrm{g}}{l}
\end{align}
with a gravitational scattering length $a_\mathrm{g}=u/(\hbar\omega)$ in analogy to the s-wave scattering length of the contact interaction.

\begin{figure}[t!]
	\includegraphics[scale=0.45]{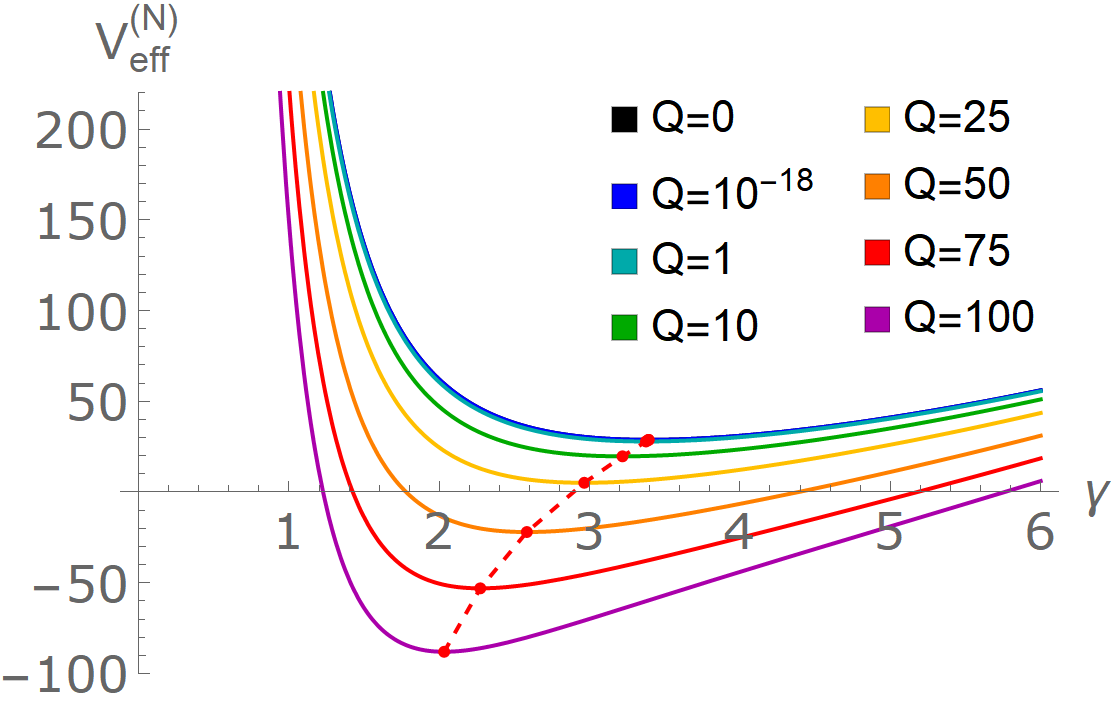}
	\caption{(Colors online) Effective potential $V_\mathrm{eff}^\mathrm{(N)}$ depending on the Gaussian width $\gamma$ for a spherically symmetric condensate and $P=446$. The curves correspond to different gravitational interaction strengths $Q$. For each curve we include the minimum as well as an interpolation given by the red dashed line.}
	\label{fig:Veff Newton}
\end{figure}

In Fig.~\ref{fig:Veff Newton} we present the effective potential of Eq.~\eqref{eq:Veff Newton} depending on the Gaussian width for various gravitational interaction strengths. Again, the minimum of each curve is marked by a red dot, and we clearly see that an increased gravitational interaction leads to smaller equilibrium widths. This is to be expected since the gravitational interaction is always attractive, while the contact interaction is here assumed to be repulsive. 

Next we take a look at the differential equations. According to the general expression~\eqref{eq:ODE general} we need to determine each spatial derivative of the Lagrangian in Eq.~\eqref{eq:LN dimless} with respect to $\gamma_j$. We now interchange the integral with the derivatives, since both variables $\gamma_j$ and $\kappa_j$ are independent. Then we first calculate the derivative with respect to $\gamma_j$. After that we apply the spherical symmetry and integrate over $\kappa_j$ in spherical coordinates. This technique allows us later on to derive the correct Hessian matrix with three eigenvalues.

The differential equation including a Newtonian interaction reads
\begin{align}\label{eq:ODE Newton sphere}
\ddot{\gamma} = -\gamma + \frac{1}{\gamma^3} + \frac{P}{\gamma^4} - \frac{Q}{\gamma^2},
\end{align}
which leads to the equilibrium cloud width given by
\begin{align}\label{eq:equi width Newton sphere}
\gamma_0^5 - \gamma_0 - P + Q\gamma_0^2 = 0.
\end{align}

\begin{figure}[t!]
	\includegraphics[scale=0.45]{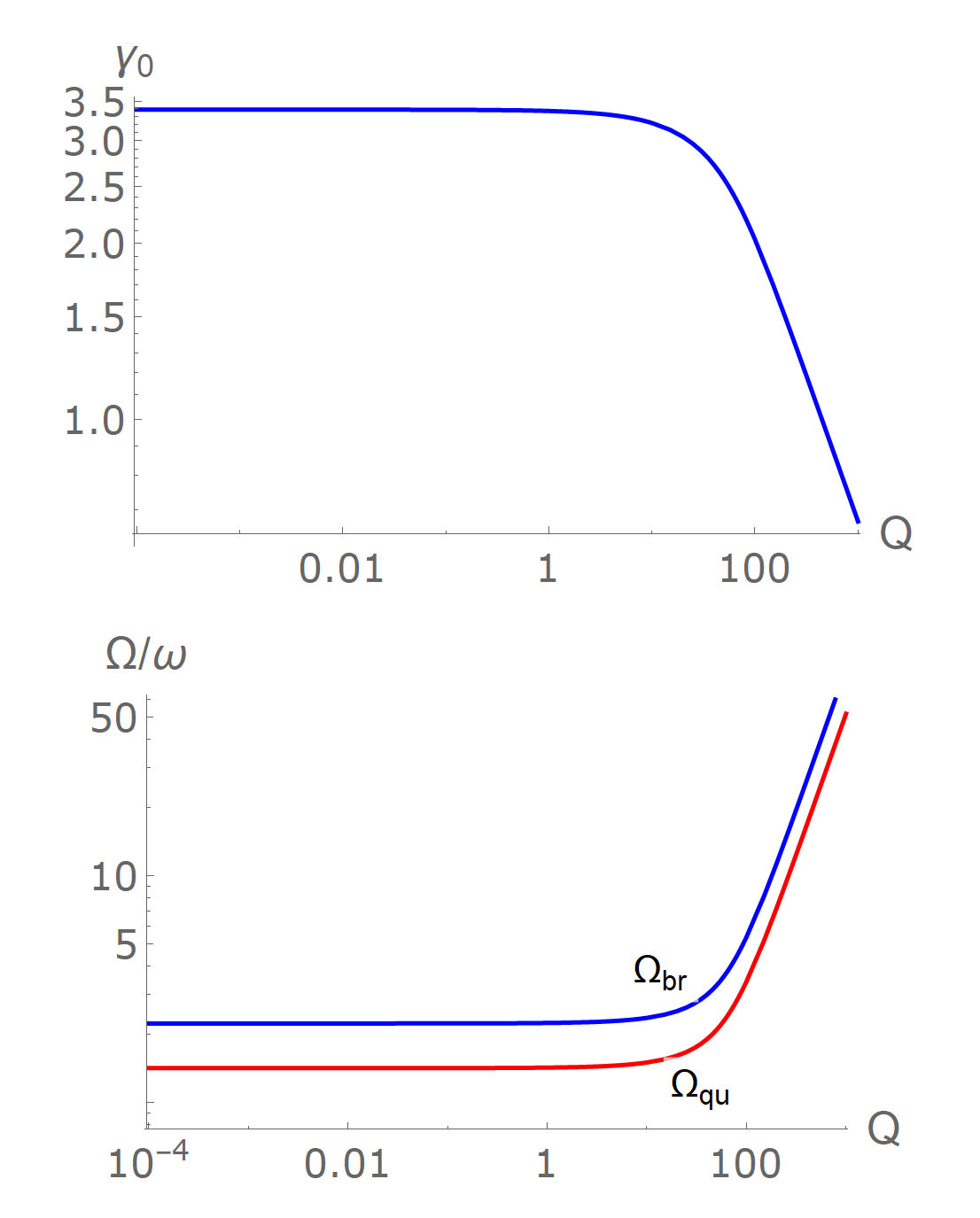}
	\caption{(Colors online) Equilibrium cloud width $\gamma_0$ and collective frequencies $\Omega_\mathrm{br}^\mathrm{(N)}$ and $\Omega_\mathrm{qu}^\mathrm{(N)}$ of a spherical condensate with $P=446$ depending on the gravitational interaction strength $Q$.}
	\label{fig:sphere Newton}
\end{figure}

Similarly, we derive the elements of the Hessian matrix for a Newtonian interaction with the second derivatives of the Lagrangian~\eqref{eq:LN dimless} according to Eq.~\eqref{eq:Mint general}. The corrections to the diagonal and off-diagonal elements are
\begin{align}\label{eq:MN sphere}
M_\mathrm{N}^{(jj)} = -\frac{4}{5}\frac{Q}{\gamma_0^3}, \\
M_\mathrm{N}^{(jk)} = -\frac{3}{5}\frac{Q}{\gamma_0^3}.
\end{align}
Finally, the collective frequencies determined by the eigenvalues of the full Hessian are given as 
\begin{align}\label{eq:coll frequ Newton sphere}
\frac{\Omega_\mathrm{br}^\mathrm{(N)}}{\omega} = \sqrt{1+\frac{3}{\gamma_0^4}+\frac{4P}{\gamma_0^5}-\frac{2Q}{\gamma_0^3}}, \\
\frac{\Omega_\mathrm{qu}^\mathrm{(N)}}{\omega} = \sqrt{1+\frac{3}{\gamma_0^4}+\frac{P}{\gamma_0^5}-\frac{Q}{5\gamma_0^3}}.
\end{align}
Again, the frequencies of the quadrupole modes are degenerate. Moreover, both frequencies are corrected by a term proportional to $\gamma_0^{-3}$ with a slight difference in the prefactor. Note that $Q$ is by definition always positive and thus represents purely attractive interactions, as expected in case of gravity. This is also visible in Fig.~\ref{fig:sphere Newton}. There the equilibrium cloud width and the collective frequencies are shown as a function of the gravitational interaction strength. The equilibrium cloud width $\gamma_0$, which represents the size of the cloud, is drastically decreased for values $Q>10$. This is clear because the attractive interaction outweighs the repulsive contact interaction, leading to a smaller size of the condensate. On the other side, the collective frequencies are larger for $Q>10$. The values for the equilibrium cloud width again agree with the minima of the effective potential for the corresponding gravitational interaction strength shown previously.

In the end of this section we mention a typical value for the gravitational interaction strength $Q$. Using the definition in Eq.~\eqref{eq:def Q sphere} and the example of a $^{87}\mathrm{Rb}$ condensate, we obtain $Q\approx 4\cdot10^{-19}$, which is about twenty orders of magnitude smaller than the contact interaction strength. Consequently, it is very unlikely that the Newtonian interaction can be measured in a laboratory. However, the method discussed here will be useful later on for the Yukawa-like interaction. With the typical value of $Q$ we calculate the equilibrium cloud width and the collective frequencies
\begin{align}\label{eq:Newton typ}
\gamma_0^\mathrm{(N)} &= 2.89\;\mathrm{\mu m}, \nonumber\\
\Omega_\mathrm{br}^\mathrm{(N)} &= 2.234\;\mathrm{kHz},\;\Omega_\mathrm{qu}^\mathrm{(N)} = 1.420\;\mathrm{kHz}
\end{align}
which serve as a reference for later calculations.

\subsection{Yukawa interaction}\label{sec:sphere Yukawa}

Analoguous to the Newtonian case, we evaluate the integrals in Eq.~\eqref{eq:LYuk dimless} with the methods discussed in the previous section.

\begin{figure}[t!]
	\includegraphics[scale=0.45]{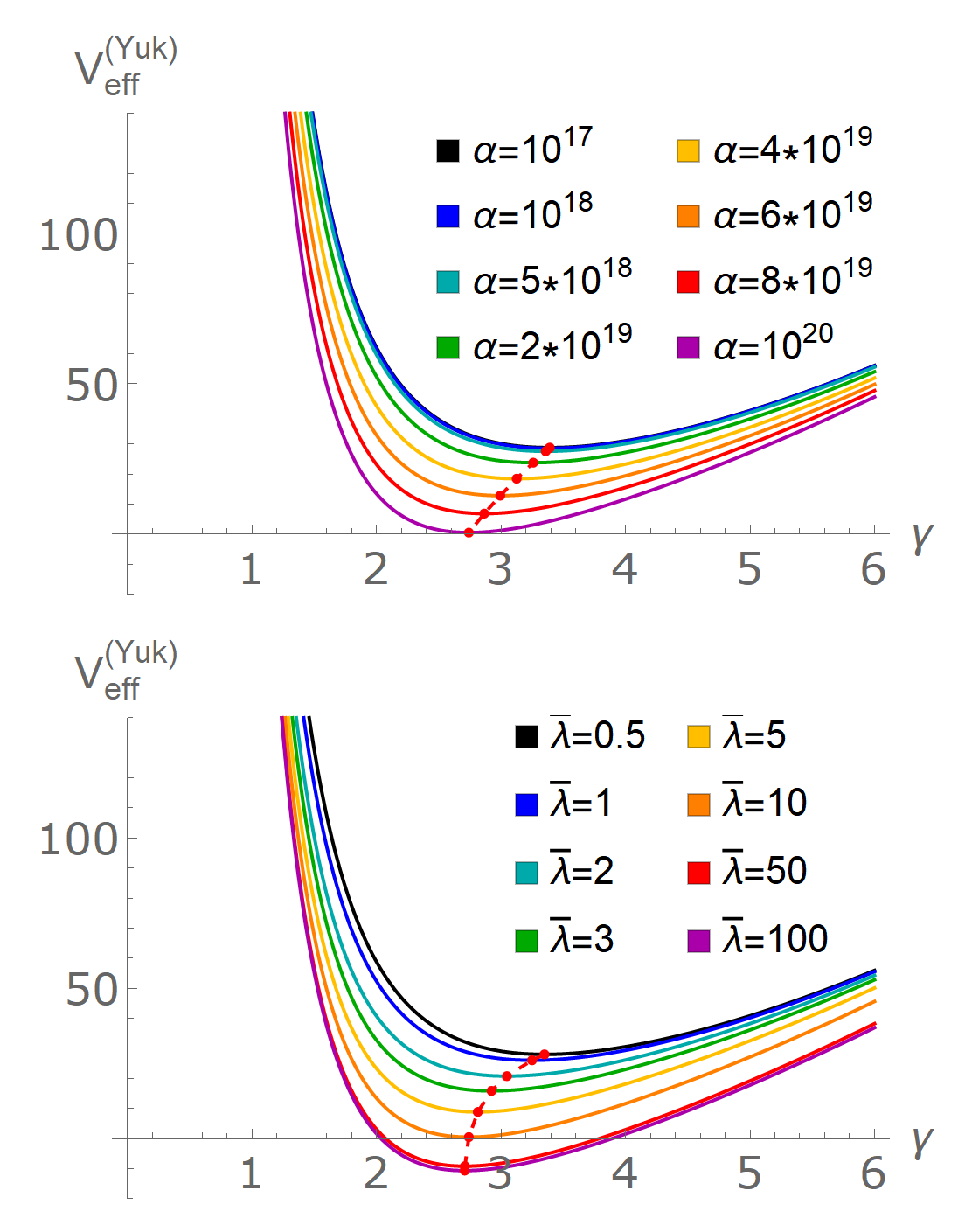}
	\caption{(Colors online) Effective potential $V_\mathrm{eff}^\mathrm{(Yuk)}$ depending on the Gaussian width $\gamma$ for a spherically symmetric condensate with Yukawa-like interaction. The red dots indicate the minimum of each curve and an interpolation is given by the red dashed line. In both pictures we choose $P=446$ and $Q=4\cdot10^{-19}$ and in addition $\bar{\lambda}=10$ in the upper panel and $\alpha=10^{20}$ in the lower picture.}
	\label{fig:Veff Yuk}
\end{figure}

In spherical coordinates the integrals, in particular the radial integral, can be looked up in the literature, e.g. Ref.~\cite{Gradsteyn}. For the effective potential the result reads
\begin{align}\label{eq:Veff Yuk}
V_\mathrm{eff}^{\mathrm{(Yuk)}} = &\frac{3}{2}\left(\gamma^2+\frac{1}{\gamma^2}\right) + \frac{P}{\gamma^3} - \frac{3Q}{\gamma} \nonumber\\
&- 3\alpha Q\left( \frac{1}{\gamma}-\sqrt{\frac{\pi}{2}}\frac{1}{\bar{\lambda}} \exp\left\{\frac{\gamma^2}{2\bar{\lambda}^2}\right\} \mathrm{erfc}\left[\frac{\gamma}{\sqrt{2}\bar{\lambda}}\right] \right).
\end{align}
Here we introduce the complementary error function $\mathrm{erfc}(x)$, which as usual is defined as the integral
\begin{align}
\mathrm{erfc}(x)=\frac{2}{\sqrt{\pi}}\int_x^\infty\mathrm{d}t\;\mathrm{e}^{-t^2}.
\end{align}
Since we now have two additional parameters, we show the dependency of the effective potential on both parameters in the two diagrams in Fig.~\ref{fig:Veff Yuk}. We choose $P=446$ and $Q=4\cdot10^{-19}$ and $\bar{\lambda}=10$ in the upper figure and $\alpha=10^{20}$ in the lower panel. Again, we mark the minimum of each curve with a red dot. The variation of the strength $\alpha$ is equivalent to the variation of the gravitational interaction $Q$, which is shown in Fig.~\ref{fig:Veff Newton} for the Newtonian interaction. Both are simple prefactors, i.e. both have the same effect, so that the minima are at smaller widths for higher values of the interaction strengths. However, a change in the effective range initially shows a significant change in the equilibrium width, but the equilibrium is almost constant for $\bar{\lambda}>10$ in this example. From a physical point of view, a particle at the very edge of the condensate might interact with a particle which is farthest from it for a certain effective range. An increase of the effective range would not have any effect, since the whole condensate is already contained within the range. Thus the equilibrium cloud width would not change for larger ranges.

The differential equations, the equilibrium cloud width, and the Hessian matrix are derived analogously to the previous section. To summarize we obtain the differential equation
\begin{align}\label{eq:ODE Yukawa sphere}
\ddot{\gamma} = &-\gamma + \frac{1}{\gamma^3} + \frac{P}{\gamma^4} - \frac{Q}{\gamma^2} \nonumber\\
&-\alpha Q \left( \frac{1}{\gamma^2}-\frac{1}{\bar{\lambda}^2}+\sqrt{\frac{\pi}{2}}\frac{\gamma}{\bar{\lambda}^3} \exp\left\{\frac{\gamma^2}{2\bar{\lambda}^2}\right\} \mathrm{erfc}\left[\frac{\gamma}{\sqrt{2}\bar{\lambda}}\right] \right)
\end{align}
and the equilibrium cloud width, which is given by
\begin{align}\label{eq:equi width Yukawa sphere}
0= &\gamma_0^5 - \gamma_0 - P + Q\gamma_0^2 \nonumber\\
&+ \alpha Q \left(\gamma_0^2 - \frac{\gamma_0^4}{\bar{\lambda}^2} + \sqrt{\frac{\pi}{2}}\frac{\gamma_0^5}{\bar{\lambda}^3}\exp\left\{\frac{\gamma_0^2}{2\bar{\lambda}^2}\right\}\mathrm{erfc}\left[\frac{\gamma_0}{\sqrt{2}\bar{\lambda}}\right]\right).
\end{align}

\begin{figure}[t!]
	\includegraphics[scale=0.45]{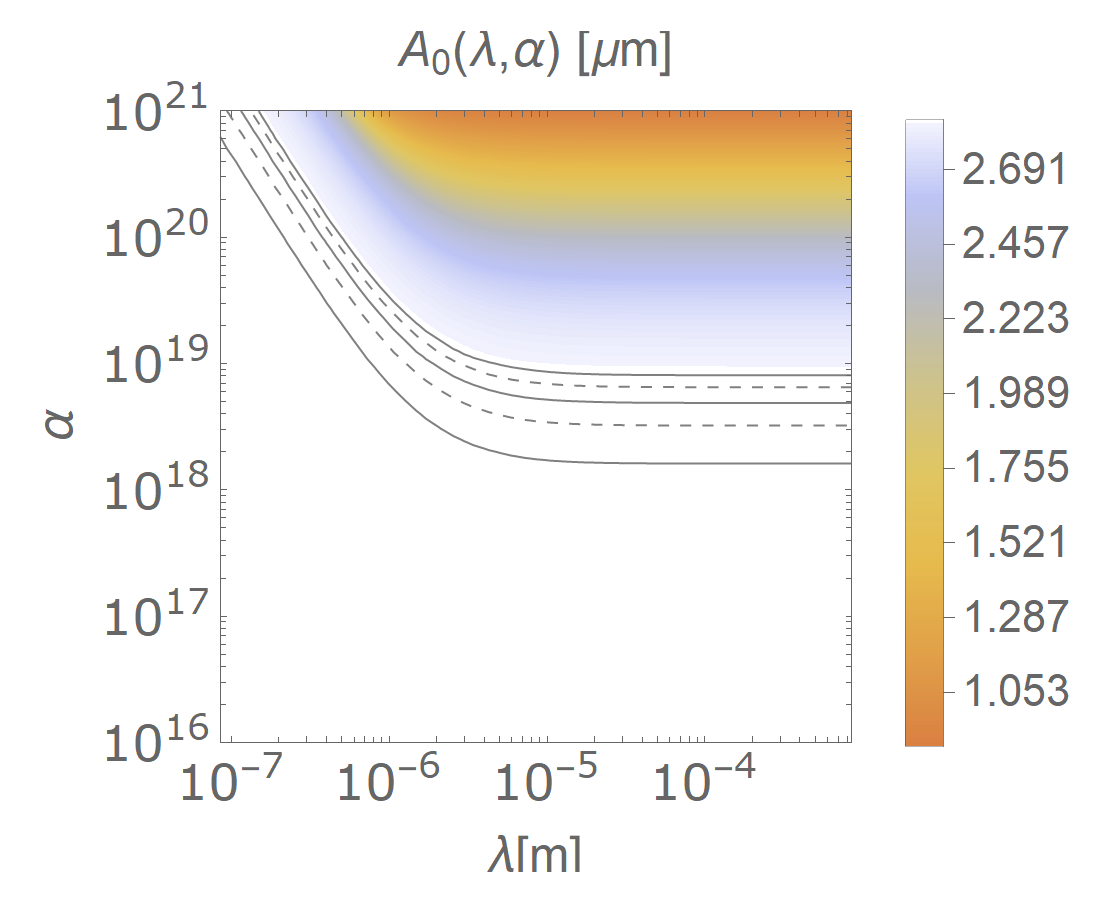}
	\caption{(Colors online) Equilibrium cloud width $A_0$ of a spherically symmetric condensate with Yukawa-like interaction as a function of the effective range $\lambda$ and the interaction strength $\alpha$. For the figure we choose $P=446$ and $Q=4\cdot10^{-19}$. The black lines show a difference of $0.01\;\mathrm{\mu m}$ to $0.05\;\mathrm{\mu m}$ to the corresponding Newtonian case given in Eq.~\eqref{eq:Newton typ}. For better visibility, the curves alternate between solid and dashed lines.}
	\label{fig:sphere Yuk width}
\end{figure}

The results of this algebraic equation are then expressed in physical units using the oscillator length $l$. Since we now have two parameters, we present in Fig.~\ref{fig:sphere Yuk width} a contour plot of the equilibrium Gaussian width $A_0$ in physical units as a function of both the effective range $\lambda$ and the interaction strength $\alpha$. In regard of a typical condensate we choose the contact and gravitational interaction strengths to be $P=446$ and $Q=4\cdot10^{-19}$. We can tell by the color code that the equilibrium width is decreased for higher values of $\alpha$ as expected. Furthermore, we also see that for $\lambda>10^{-5}\;\mathrm{m}$ the equilibrium width is independent of the effective range. This coincides with our considerations regarding Fig.~\ref{fig:Veff Yuk}. In addition to the physical values of $A_0$ we include a correction of $0.01\;\mathrm{\mu m}$ to $0.05\;\mathrm{\mu m}$ towards the equilibrium cloud width in the Newtonian interaction. These curves are marked with the black and black dashed lines. So if there exist contributions of non-Newtonian gravity in form of a Yukawa potential, we would see in an experiment an error of $0.01\;\mathrm{\mu m}$ in case of the lowest black line for the respective pair of $\alpha$ and $\lambda$. If there are no correction within a precision of $0.01\;\mathrm{\mu m}$, then all pairs of values above this curve are excluded.

The steady state and the second derivatives of the Lagrangian~\eqref{eq:LYuk dimless} then leads the diagonal and off-diagonal elements of the Hessian matrix
\begin{align}\label{eq:MYuk}
M^{(jj)} = &\;1 + \frac{3}{\gamma_0^4} + \frac{2P}{\gamma_0^5} - \frac{4}{5}\frac{Q}{\gamma_0^3} \nonumber\\
&-\alpha Q \Bigg[ \frac{4}{5}\frac{1}{\gamma_0^3} + \frac{2}{5}\frac{1}{\gamma_0\bar{\lambda}^2} + \frac{3}{5}\frac{\gamma_0}{\bar{\lambda}^4} \nonumber\\
&-\sqrt{\frac{\pi}{2}} \left( \frac{1}{\bar{\lambda}^3} + \frac{3}{5}\frac{\gamma_0^2}{\bar{\lambda}^5} \right) \exp\left\{\frac{\gamma_0^2}{2\bar{\lambda^2}}\right\} \mathrm{erfc}\left[\frac{\gamma_0}{\sqrt{2}\bar{\lambda}}\right] \Bigg], \\
M^{(jk)} = &\frac{P}{\gamma_0^5} - \frac{3}{5}\frac{Q}{\gamma_0^3} -\alpha Q \Bigg[ \frac{3}{5}\frac{1}{\gamma_0^3} - \frac{1}{5}\frac{1}{\gamma_0\bar{\lambda^2}} + \frac{1}{5}\frac{\gamma_0}{\bar{\lambda}^4} \nonumber\\
&- \sqrt{\frac{\pi}{2}}\frac{1}{5}\frac{\gamma_0^2}{\bar{\lambda^5}} \exp\left\{\frac{\gamma_0^2}{2\bar{\lambda^2}}\right\} \mathrm{erfc}\left[\frac{\gamma_0}{\sqrt{2}\bar{\lambda}}\right] \Bigg],
\end{align}
which includes the contributions of the external trap, the kinetic energy, the two-particle contact interaction as well as the Newtonian and Yukawa-like two-particle interaction. The eigenvalues and the collective frequencies are calculated numerically.

\begin{figure}[t!]
	\includegraphics[scale=0.45]{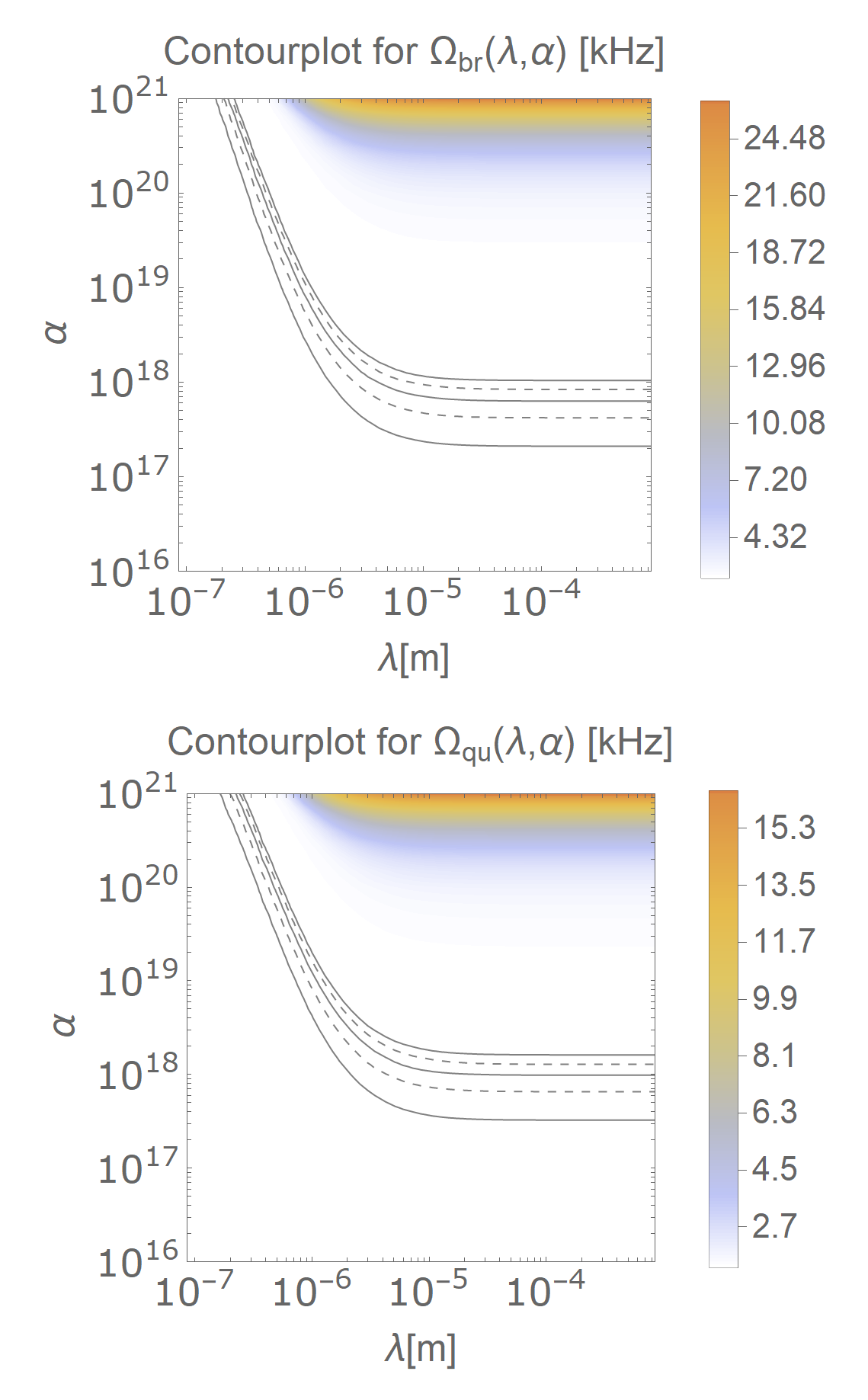}
	\caption{(Colors online) Collective frequencies $\Omega_\mathrm{br}$ and $\Omega_\mathrm{qu}$ of a spherically symmetric condensate with Yukawa-like interaction depending on the the effective range $\lambda$ and the interaction strength $\alpha$. For the figure we choose $P=446$ and $Q=4\cdot10^{-19}$. The black lines show a difference of $1\;\mathrm{Hz}$ to $5\;\mathrm{Hz}$ to the corresponding Newtonian case given in Eq.~\eqref{eq:Newton typ}. For better visibility, the curves alternate between solid and dashed lines.}
	\label{fig:sphere Yuk frequ}
\end{figure}

If we are only interested in the radial oscillation, we can also calculate the second derivative of the effective potential and divide by three as we only take into account one radial direction instead of three dimensions. The result is a analytical expression for the frequency of the breathing mode
\begin{align}\label{eq: breathing frequency analytic Yuk sphere}
\left(\frac{\Omega_\mathrm{br}}{\omega}\right)^2 &=\;1 + \frac{3}{\gamma_0^4} + \frac{4P}{\gamma_0^5} - \frac{2Q}{\gamma_0^3} - \alpha Q \Bigg[ \frac{2}{\gamma_0^3} + \frac{\gamma_0}{\bar{\lambda}^4} \nonumber\\
&- \sqrt{\frac{\pi}{2}}\left( \frac{1}{\bar{\lambda}^3} + \frac{\gamma_0^2}{\bar{\lambda^5}} \right) \exp\left\{\frac{\gamma_0^2}{2\bar{\lambda}^2}\right\}\mathrm{erfc}\left[\frac{\gamma_0}{\sqrt{2}\bar{\lambda}}\right] \Bigg].
\end{align}
However, in one radial direction it is not possible to derive the quadrupole modes in that way.

\begin{figure}[t!]
	\includegraphics[scale=0.45]{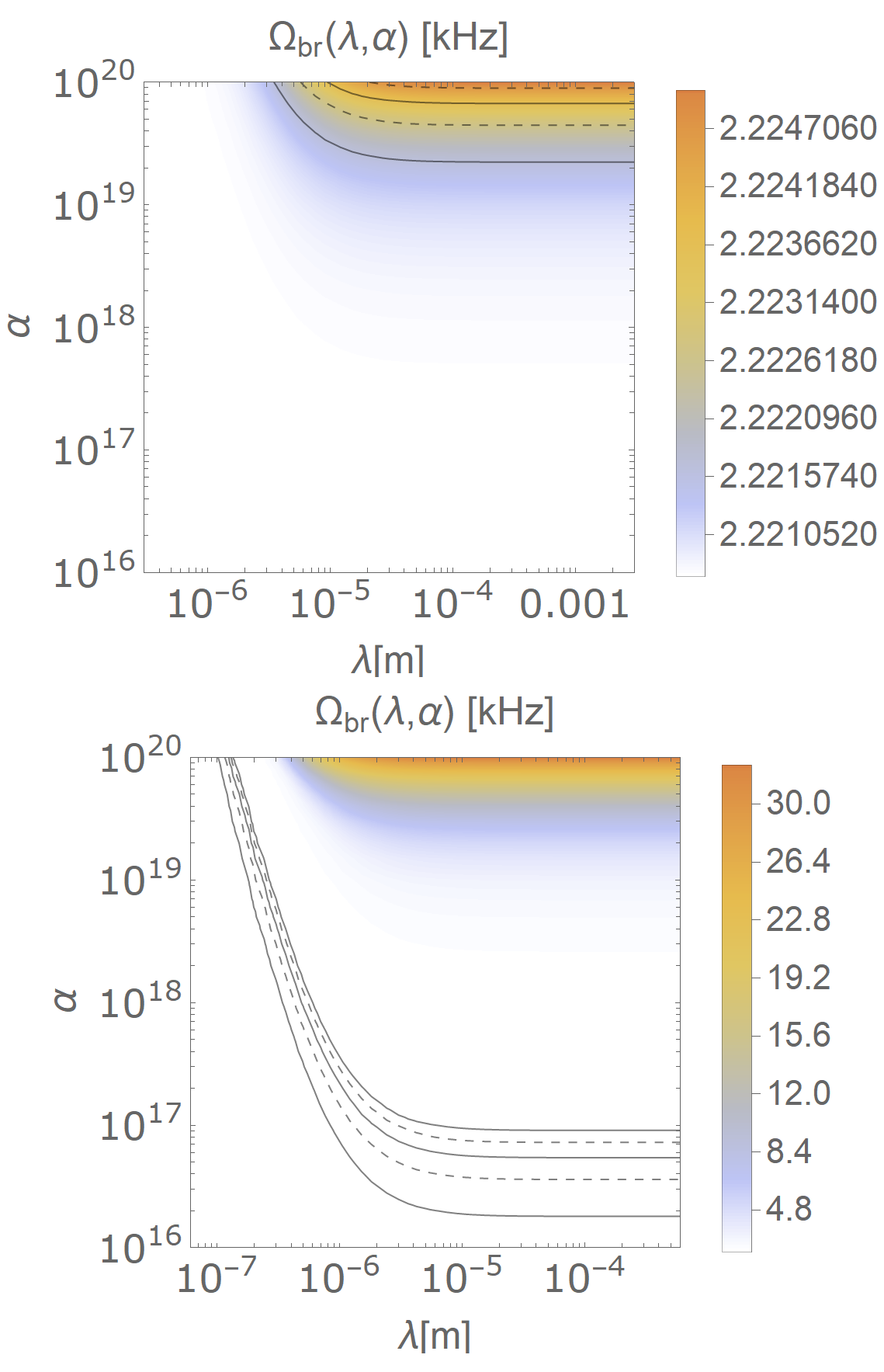}
	\caption{(Colors online) Collective frequency $\Omega_\mathrm{br}$ of the breathing mode for varying mass $m$. In the upper panel we have $^{7}\mathrm{Li}$ and in the lower plot $^{174}\mathrm{Yb}$. To exclude the dependency of the s-wave scattering we set $a_\mathrm{s}=10^{-9}\;\mathrm{m}$. The black lines show a difference of $1\;\mathrm{Hz}$ to $5\;\mathrm{Hz}$ to the corresponding Newtonian case given in Eq.~\eqref{eq:Newton typ}. For better visibility, the curves alternate between solid and dashed lines.}
	\label{fig:sphere Yuk frequ m}
\end{figure}

Analogous to the equilibrium cloud width we show in Fig.~\ref{fig:sphere Yuk frequ} both collective frequencies of the breathing mode and the quadrupole modes. Again, the frequencies are increased for a larger interaction strength, and due to the equilibrium cloud width they are independent of the effective range for $\lambda>10^{-5}\;\mathrm{m}$ in this example. The color code also reveals that the frequency of the breathing mode is generally larger than that of the quadrupole modes. The black and black dashed lines here show a difference of $1\;\mathrm{Hz}$ to $5\;\mathrm{Hz}$ compared to the collective frequencies only including a Newtonian interaction. If we compare both collective frequencies, we see that the frequency corresponding to the breathing mode leads to better constraints on both Yukawa parameters.

\begin{figure}[t!]
	\includegraphics[scale=0.45]{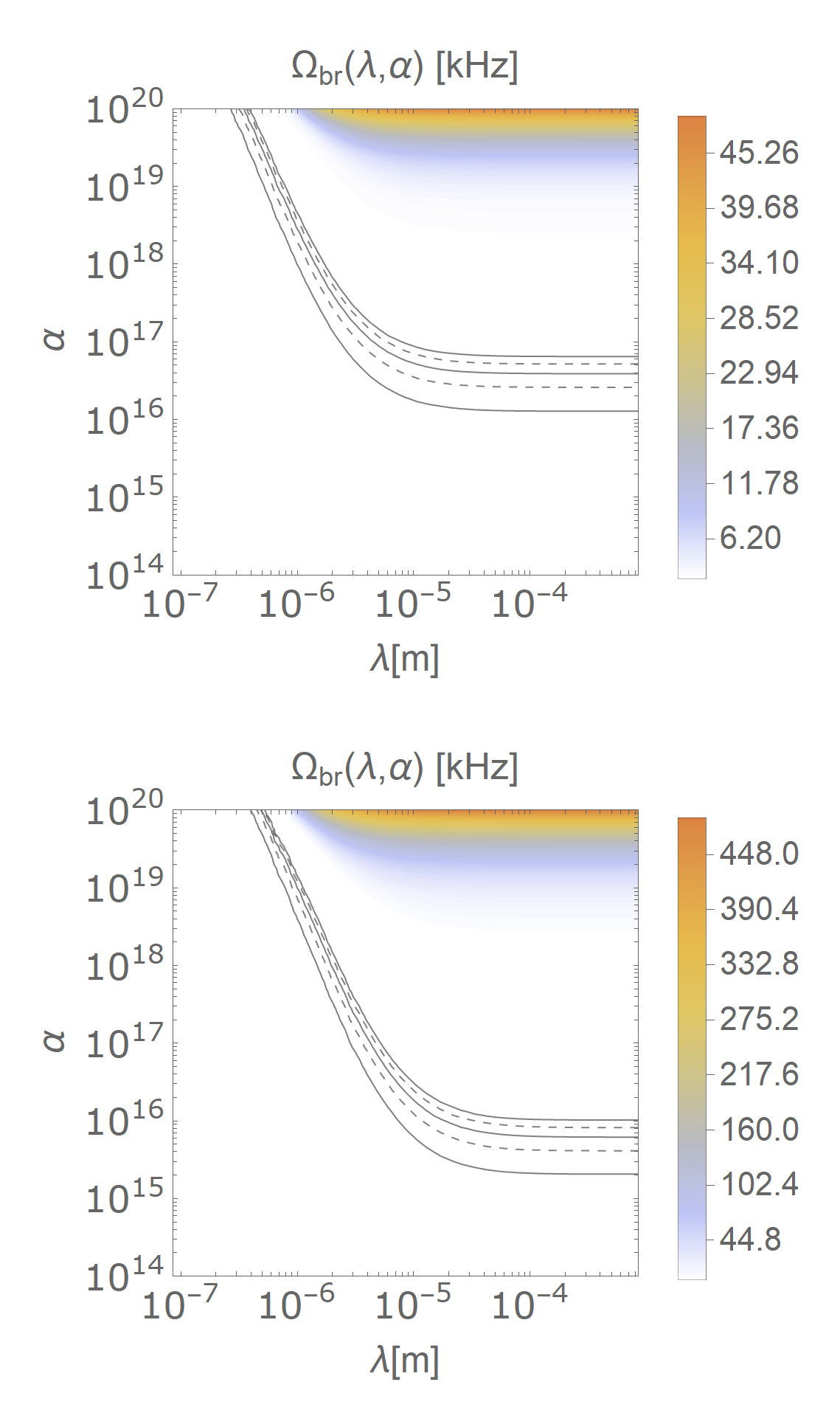}
	\caption{(Colors online) Collective frequency $\Omega_\mathrm{br}$ of the breathing mode for varying particle number $N$. We consider $N=10^7$ (top) and $N=10^9$ (bottom). The black lines show a difference of $1\;\mathrm{Hz}$ to $5\;\mathrm{Hz}$ to the corresponding Newtonian case given in Eq.~\eqref{eq:Newton typ}. For better visibility, the curves alternate between solid and dashed lines.}
	\label{fig:sphere Yuk frequ N}
\end{figure}

In the following we present the influence of experimentally accessible parameter. Since the results of both frequencies are quite similar, but the breathing frequency leads to slightly better constraints, we only consider here the collective frequency of the breathing mode. In the numerical calculation we now change one of the following values: the mass of the atomic species, the particle number, the s-wave scattering length, and the trap frequency. Note that a change of any of these values affects the contact interaction strength $P$ and the gravitational interaction strength $Q$ given by their definition~\eqref{eq:def P} and~\eqref{eq:def Q sphere}, respectively. As the basis we take the values used in Fig.~\ref{fig:sphere Yuk frequ}, namely a $^{87}\mathrm{Rb}$ condensate with $N=10^5$ particles, the scattering length $a_\mathrm{s}=90a_0$, and the trap frequency $\omega=1\;\mathrm{kHz}$.

In the context of gravity, the most obvious choice to increase the interaction is to increase the mass of the atomic species used to create the condensate. In Fig.~\ref{fig:sphere Yuk frequ m} we show the results for a $^{7}\mathrm{Li}$ and a $^{174}\mathrm{Yb}$ as the lightest and heaviest species condensed so far~\cite{Bradley, Bradley2, Takahashi}. For a better comparison we set the s-wave scattering length fixed at $a_\mathrm{s}=1\;\mathrm{nm}$, although each atomic species differs in that length. In an experiment this might be addressed by a Feshbach resonance. As expected, the constraints for the lighter atomic species are worse than for the typical $^{87}\mathrm{Rb}$ condensate shown previously. On the other side, the heavier candidate improves the constraints by roughly one order of magnitude for the interaction strength $\alpha$ and very slightly for the effective range $\lambda$.

Next we increase the particle number. As we see in Fig.~\ref{fig:sphere Yuk frequ N}, the lowest line for a hypothetical particle number $N=10^9$ would reach a value of the order $\alpha=10^{15}$. Noticeably, the frequency is drastically increased reaching hundreds of $\mathrm{kHz}$ at the upper right panel of the figure.

\begin{figure}[t!]
	\includegraphics[scale=0.45]{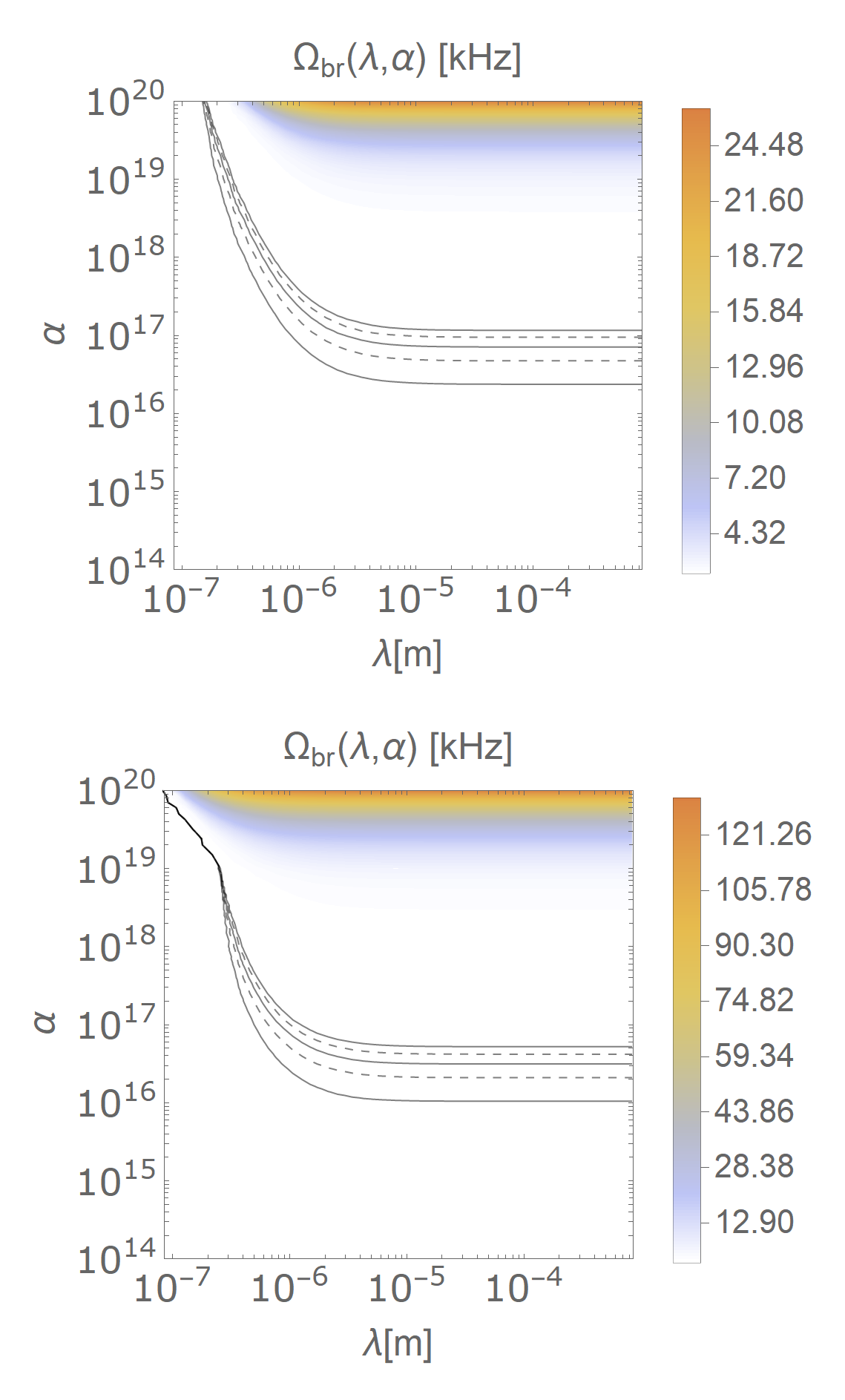}
	\caption{(Colors online) Collective frequency $\Omega_\mathrm{br}$ of the breathing mode for varying s-wave scattering length $a_\mathrm{s}$. We consider $a_\mathrm{s}=10^{-10}\mathrm{m}$ (top) and $a_\mathrm{s}=10^{-11}\mathrm{m}$ (bottom). The black lines show a difference of $1\;\mathrm{Hz}$ to $5\;\mathrm{Hz}$ to the corresponding Newtonian case given in Eq.~\eqref{eq:Newton typ}. For better visibility, the curves alternate between solid and dashed lines.}
	\label{fig:sphere Yuk frequ a}
\end{figure}

The s-wave scattering length can be modified by Feshbach resonances, and Fig.~\ref{fig:sphere Yuk frequ a} shows that smaller values of $a_\mathrm{s}$ are favorable in regard of constraints. The frequencies are increased for smaller s-wave scattering lengths.

\begin{figure}[t!]
	\includegraphics[scale=0.45]{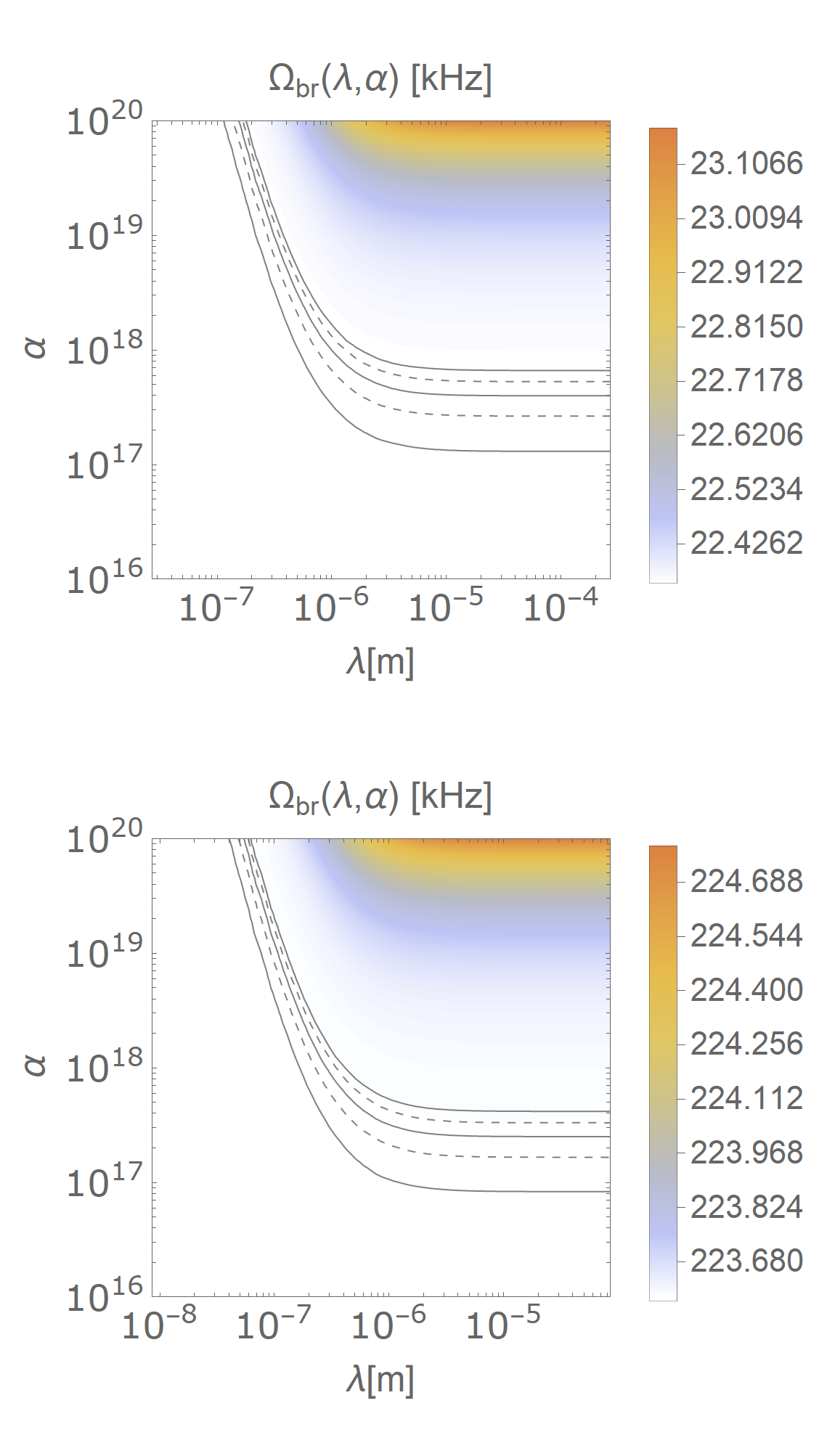}
	\caption{(Colors online) Collective frequency $\Omega_\mathrm{br}$ of the breathing mode for varying trap frequency $\omega$. We choose as examples $\omega=10\;\mathrm{kHz}$ (top) and $\omega=100\;\mathrm{kHz}$ (bottom). The black lines show a difference of $1\;\mathrm{Hz}$ to $5\;\mathrm{Hz}$ to the corresponding Newtonian case given in Eq.~\eqref{eq:Newton typ}. For better visibility, the curves alternate between solid and dashed lines.}
	\label{fig:sphere Yuk frequ omega}
\end{figure}

Finally, we investigate the influence of the trap frequency. The results are presented in Fig.~\ref{fig:sphere Yuk frequ omega}, which shows better constraints for higher trapping frequencies. Again, increasing the trapping frequency leads to a larger collective frequency.

\begin{figure}[t!]
	\includegraphics[scale=1.5]{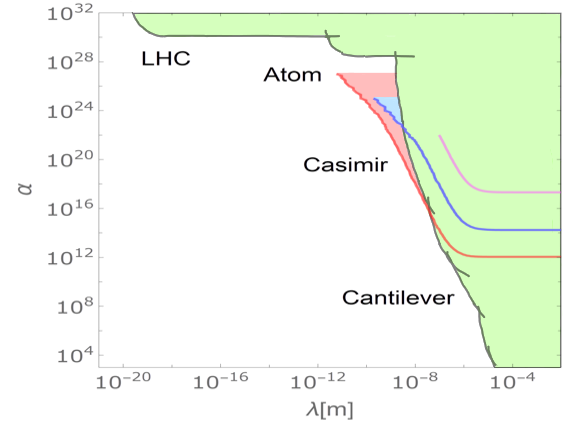}
	\caption{(Colors online) Comparison between the experimental data taken from Ref.~\cite{Murata} and our theoretical model for the constraints of the Yukawa parameters. The green shaded area is excluded with a confidence level of 95\%. The magenta line is the result for a typical $^{87}\mathrm{Rb}$ condensate ($N=10^5$, $a_\mathrm{s}=90\;a_0$, $\omega=1\;\mathrm{kHz}$), the blue line for realizable $^{174}\mathrm{Yb}$ condensate ($N=10^8$, $a_\mathrm{s}=10^{-10}\;\mathrm{m}$, $\omega=10\;\mathrm{kHz}$) and the red line for a hypothetical $^{174}\mathrm{Yb}$ condensate ($N=10^{11}$, $a_\mathrm{s}=10^{-11}\;\mathrm{m}$, $\omega=10^3\;\mathrm{kHz}$). For all curves we assume a frequency correction of $1\;\mathrm{Hz}$ compared to the Newtonian case given in Eq.~\eqref{eq:Newton typ}. The blue shaded and the red shaded area indicate an improvement of the constraints.}
	\label{fig:sphere Yuk frequ comp}
\end{figure}

To summarize, the constraints found by a typical $^{87}\mathrm{Rb}$ condensate, shown in Fig.~\ref{fig:sphere Yuk frequ}, can be improved by using a $^{174}\mathrm{Yb}$ condensate with larger particle number, smaller s-wave scattering length and higher trapping frequency.
In the end of this section we compare our results to experimentally verified data in context of constraints for the Yukawa parameters, which are explained in detail in Ref.~\cite{Murata}. In Fig.~\ref{fig:sphere Yuk frequ comp} we show the experimental results with our results of three different condensates side by side. If one measures the collective frequency of the breathing mode in a typical $^{87}\mathrm{Rb}$ condensate with an accuracy of $1\;\mathrm{Hz}$, the constraints, in particular for the effective range $\lambda$, are surprisingly close to experimental data, although off roughly one order of magnitude. A realizable $^{174}\mathrm{Yb}$ condensate could almost confirm the data as shown with the blue line. A hypothetical $^{174}\mathrm{Yb}$ with $N=10^{11}$ particles and a s-wave scattering length $a_\mathrm{s}=10^{-11}\;\mathrm{m}$ trapped within a harmonic potential with a frequency $\omega=10^8\;\mathrm{Hz}$ could in principle lead to significant improvements.

\section{Cylindrical condensates}\label{sec:axial}

In this section we generalize our results so far to axially symmetric condensates. First of all, these condensates are more realistic in the experiment, since creating a perfect sphere is usually quite challenging. There are also studies about a dimensional reduction of the three dimensional Gross-Pitaevksii equation to effective lower dimensions, see for example Ref.~\cite{Salasnich}, and it has been shown that the self-interaction factor $g$ in the contact interaction is enhanced in lower dimensions~\cite{Petrov, Olshanii}. So this might be a possibility to further constrain the Yukawa parameters.

In the following we apply a cylindrical symmetry to the expressions derived in Sec.~\ref{sec:general}. For this, we choose two dimensions to be equal, i.e. $\nu_x=\nu_y=\nu_a$. The same applies to the Gaussian widths. Additionally, we define the trap aspect ratio $\zeta=\nu_z/\nu_a$ as the ratio between the frequencies in the longitudinal and transversal direction. If we now set $\nu_a=1$, we can easily distinguish two cases: i) $\zeta<1$ describes a cigar-shaped condensate and ii) $\zeta>1$ leads to a disk-shaped configuration. The limit $\zeta=1$ then corresponds to a spherical condensate discussed in the previous section. 

\subsection{Contact interaction}\label{sec:axial con}

For a condensate with contact interaction we have already derived a general expression for a set of differential equations in any symmetry, which are given in Eq.~\eqref{eq:ODE con}. Consequently, in the cylindrical symmetry these reduce to two differential equations for the cloud widths $\gamma_a$ and $\gamma_z$. Analogous to the previous section we define the equilibrium, such that the accelerations vanish, thus
\begin{align}
&\mathrm{i)}\;\;\gamma_{a0}^5 - \gamma_{a0} - P \frac{\gamma_{a0}}{\gamma_{z0}} =0, \label{eq:equi width con axial 1}\\
&\mathrm{ii)}\;\;\zeta^2\gamma_{z0}^5 - \gamma_{z0} - P \frac{\gamma_{z0}^2}{\gamma_{a0}^2} =0. \label{eq:equi width con axial 2}
\end{align}
Due to the particle interaction this is a system of two coupled equations, which have to be solved simultaneously. Again, the equilibrium is used in the corresponding Hessian matrix to calculate the eigenvalues numerically, which analogously leads to the collective frequencies. The eigenvectors give insights in the modes containing a breathing mode and two distinguishable quadrupole modes, the radial quadrupole mode and the out-of-phase quadrupole mode, also discussed in Ref.~\cite{Perez}.

\begin{figure}[t!]
	\includegraphics[scale=0.45]{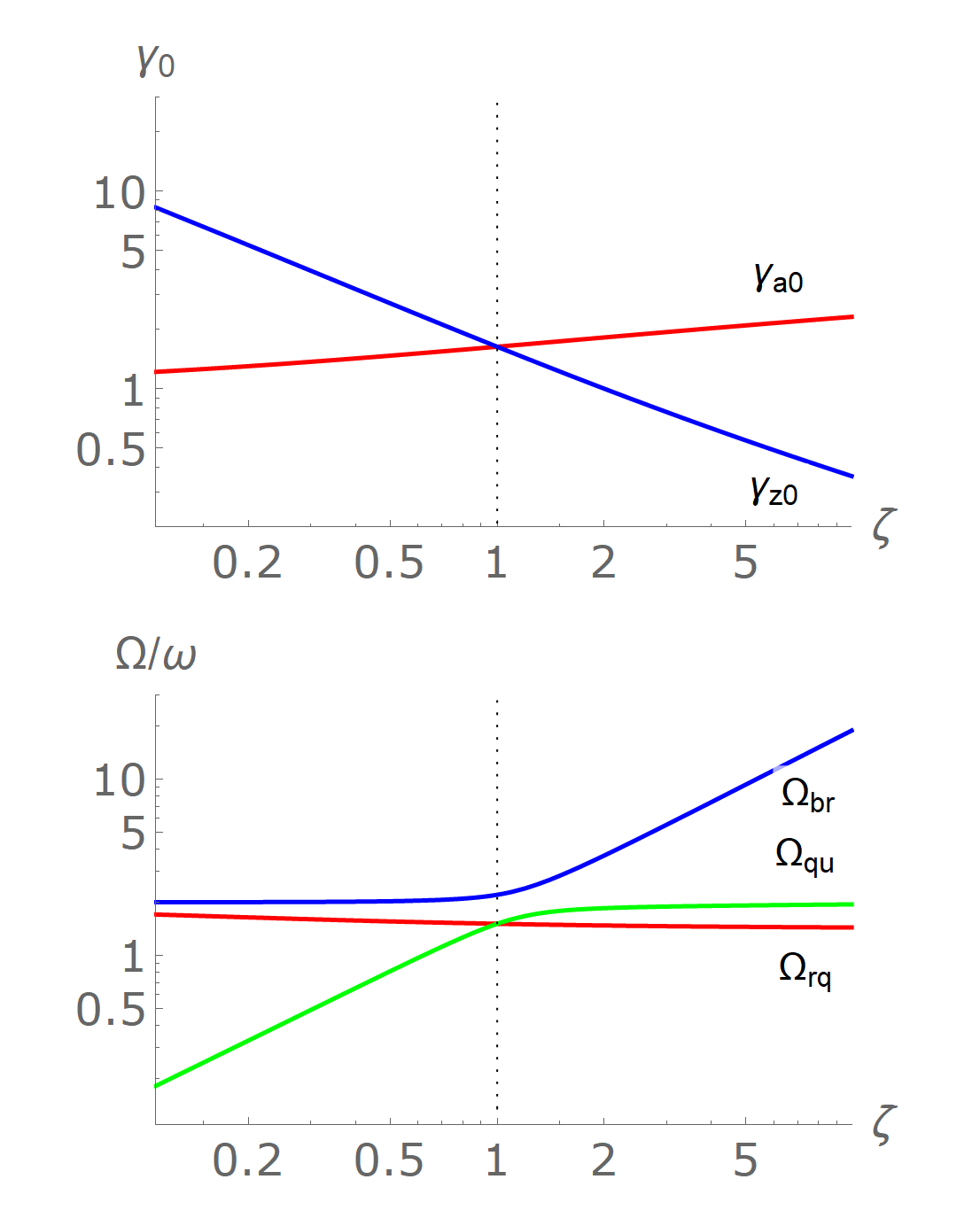}
	\caption{(Colors online) Equilibrium cloud widths $\gamma_{a0}$ and $\gamma_{z0}$, and collective frequencies $\Omega_\mathrm{br}/\omega$, $\Omega_\mathrm{qu}/\omega$, and $\Omega_\mathrm{rq}/\omega$ for an axially symmetric condensate as a function of the trap aspect ratio $\zeta$. The contact interaction strength is $P=446$. The black dotted line indicates the spherical case, on its left side the condensate is cigar-shaped, on the right side disk-shaped.}
	\label{fig:axial con aspect}
\end{figure}

As we have now two additional parameters, the aspect ratio $\zeta$ and the contact interaction strength $P$, we divide the discussion in two parts. First we show in Fig.~\ref{fig:axial con aspect} the results depending on the aspect ratio, while the interaction strength is set fixed at $P=446$. Based on the cloud widths we can directly read off the shape of the condensate. For $\zeta<1$ the width in the transversal direction $\gamma_{z0}$ is always larger than $\gamma_{a0}$, indicating a cigar-shaped form. For $\zeta>1$ it is the opposite, and at the point $\zeta=1$ both equilibrium cloud widths coincide which indicates the spherical case as mentioned before. There the value of $\gamma_0\approx 3.47$ is identical to the result in the previous section at $P=446$.

With the steady-state equations~\eqref{eq:equi width con axial 1} and~\eqref{eq:equi width con axial 2} we also mention two limiting cases. For non-interacting particles the solutions are $\gamma_{a0}=1$ and $\gamma_{z0}=\sqrt{\zeta^{-1}}$, while in the Thomas-Fermi limit we get $\gamma_{a0}^5=P\zeta$ and $\gamma_{z0}^5=P\zeta^{-4}$. Both cases show an exponential law, which resembles straight lines with different slopes in the double logarithmic scale of Fig.~\ref{fig:axial con aspect}.

In Fig.~\ref{fig:axial con aspect} also the collective frequencies are shown as a function of the aspect ratio. We now have three distinguishable frequencies except for $\zeta=1$, where both frequencies of the quadrupole modes are degenerate. Again, this fact and the exact values are identical to those derived in the previous section for spherical condensates.

\begin{figure}[t!]
	\includegraphics[scale=0.45]{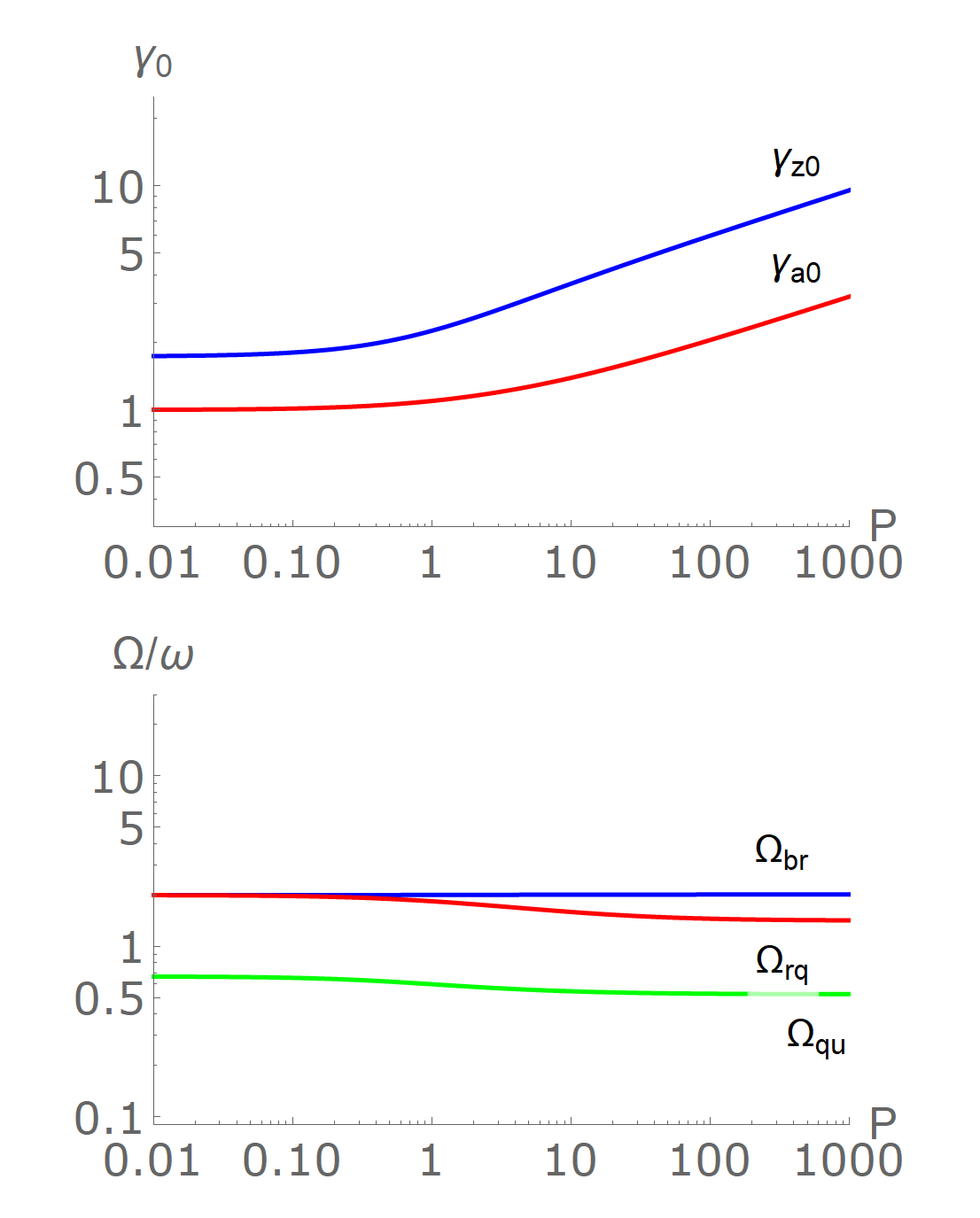}
	\caption{(Colors online) Equilibrium cloud widths $\gamma_{a0}$ and $\gamma_{z0}$, and collective frequencies $\Omega_\mathrm{br}/\omega$, $\Omega_\mathrm{qu}/\omega$, and $\Omega_\mathrm{rq}/\omega$ for a cigar-shaped condensate with a aspect ratio $\zeta=1/7$ depending on the contact interaction strength $P$.}
	\label{fig:axial con strength cigar}
\end{figure}

\begin{figure}[t!]
	\includegraphics[scale=0.45]{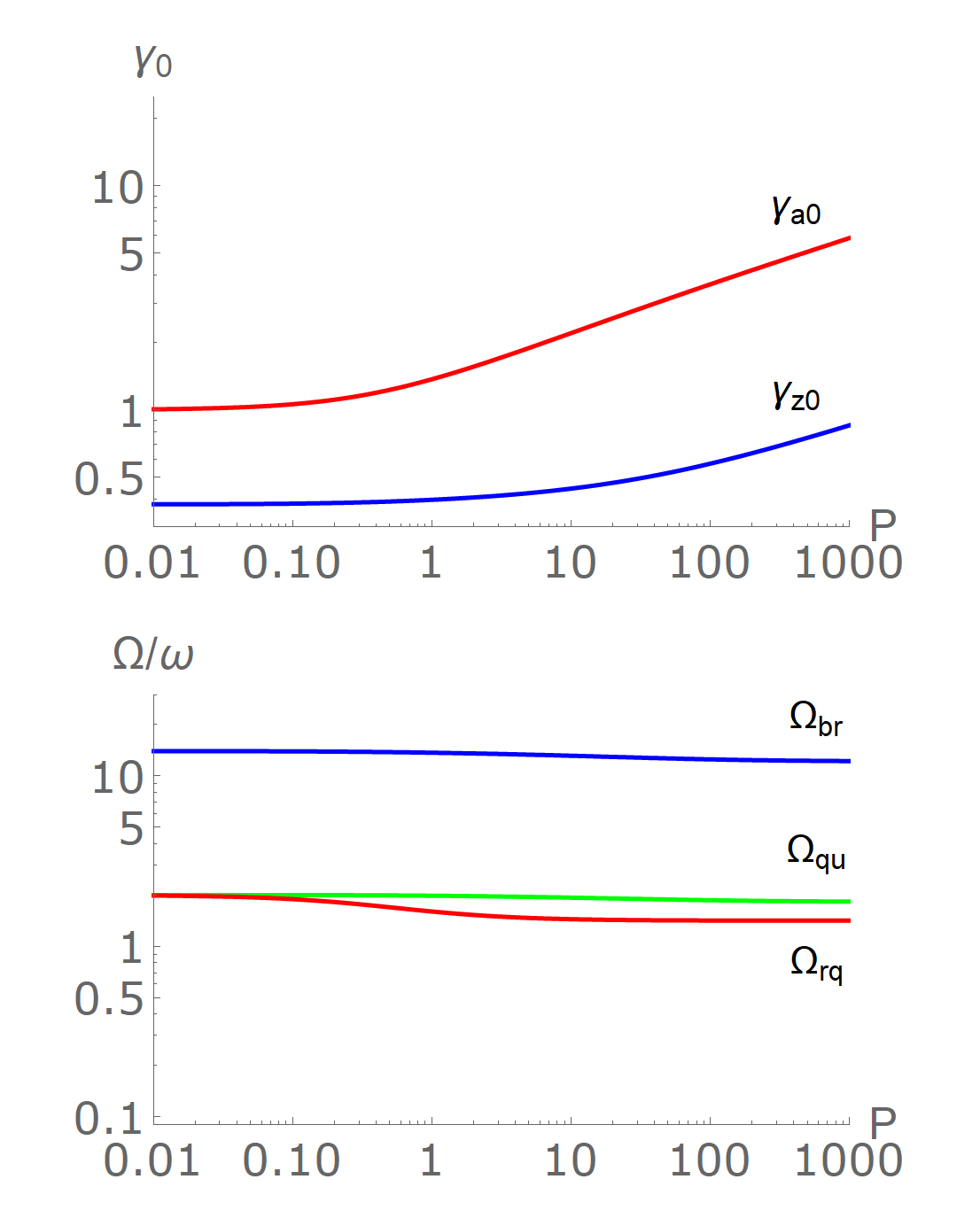}
	\caption{(Colors online) Equilibrium cloud widths $\gamma_{a0}$ and $\gamma_{z0}$, and collective frequencies $\Omega_\mathrm{br}/\omega$, $\Omega_\mathrm{qu}/\omega$, and $\Omega_\mathrm{rq}/\omega$ for a disk-shaped condensate with a aspect ratio $\zeta=7$ depending on the contact interaction strength $P$.}
	\label{fig:axial con strength disk}
\end{figure}

Next, for reasons of completeness we present the dependency of the equilibrium cloud widths and the collective frequencies on the contact interaction strength $P$. As two examples we show a cigar-shaped condensate with the aspect ratio $\zeta=1/7$ in Fig.~\ref{fig:axial con strength cigar} and a disk-shaped configuration with $\zeta=7$ in Fig.~\ref{fig:axial con strength disk}. The individual values at $P=446$ can be crosschecked with Fig.~\ref{fig:axial con aspect} for the corresponding aspect ratios. Other than that, the curves look qualitatively similar to those shown in Fig.~\ref{fig:sphere con}, only that we have now two individual cloud widths and two distinguishable quadrupole frequencies.

\subsection{Newtonian interaction}\label{sec:axial Newton}

In case of a cylindrical condensate with a Newtonian particle interaction we calculate the first derivative of the Lagrangian $L_\mathrm{N}$ in Eq.~\eqref{eq:LN dimless} with respect to $\gamma_j$ and then insert the cylindrical coordinates for the integration. After the integration over the polar coordinates $\varphi$ and $\kappa_\rho$ we formally define the function
\begin{align}\label{eq:KN axial def}
K_\mathrm{N} = \frac{1}{2}\int_{-\infty}^\infty\mathrm{d}\kappa_z\;\Gamma\left(0,\frac{1}{2}\gamma_a^2\kappa_z^2\right)\mathrm{e}^{\frac{1}{2}\gamma_a^2\kappa_z^2}\exp\left\{-\frac{1}{2}\gamma_z^2\kappa_z^2\right\}
\end{align}
as a preparation for the Yukawa-like interaction. Here $\Gamma(0,\gamma_a^2\kappa_z^2/2)$ denotes the incomplete gamma function. However, the integral over $\kappa_z$ can be solved analytically in the special case of the Newtonian interaction. The result is
\begin{align}\label{eq:KN axial result}
K_\mathrm{N} = \frac{\sqrt{2\pi}}{\gamma_a}\frac{\mathrm{arcsinh}\sqrt{\frac{\gamma_z^2}{\gamma_a^2}-1}}{\sqrt{\frac{\gamma_z^2}{\gamma_a^2}-1}}.
\end{align}
Note that the function $K_\mathrm{N}$ by its definition in Eq.~\eqref{eq:KN axial def} is related to a gravitational interaction term in an effective one dimensional Gross-Pitaevskii equation. The method of the dimensional reduction of the Gross-Pitaevksii equation with contact interaction is described in Ref.~\cite{Salasnich}.

However, with the solution~\eqref{eq:KN axial result} we now derive two differential equations for the Gaussian widths $\gamma_a$ and $\gamma_z$, namely
\begin{align}
&\mathrm{i)}\;\; \ddot{\gamma}_a = -\gamma_a + \frac{1}{\gamma_a^3} + \frac{P}{\gamma_z}\frac{1}{\gamma_a^3} + \frac{3 Q}{2\sqrt{2\pi}}\;\partial_{\gamma_a} K_\mathrm{N}, \\
&\mathrm{ii)}\;\; \ddot{\gamma}_z = -\zeta^2\gamma_z + \frac{1}{\gamma_z^3} + \frac{P}{\gamma_a^2}\frac{1}{\gamma_z^2} + \frac{3 Q}{\sqrt{2\pi}}\;\partial_{\gamma_z}  K_\mathrm{N}.
\end{align}
For the sake of simplicity, we do not write down the derivatives explicitly but please do note that the derivatives differ in both equations. Furthermore, it can be shown that in the limit $\gamma_a\rightarrow\gamma_z$ both differential equations indeed reduce to the equation given in Eq.~\eqref{eq:ODE Newton sphere} for the spherical case.

The steady state is then determined by
\begin{align}
&\mathrm{i)}\;\; \gamma_{a0}^5 - \gamma_{a0} - P\frac{\gamma_{a0}}{\gamma_{z0}}- \gamma_{a0}^4 \frac{3 Q}{2\sqrt{2\pi}}\;\left.\partial_{\gamma_a} K_\mathrm{N}\right\vert_{\boldsymbol{\gamma}=\boldsymbol{\gamma}_0} = 0, \\
&\mathrm{ii)}\;\; \zeta^2\gamma_{z0}^5 - \gamma_{z0} - P\frac{\gamma_{z0}^2}{\gamma_{a0}^2} - \gamma_{z0}^4 \frac{3 Q}{\sqrt{2\pi}}\;\left.\partial_{\gamma_z}  K_\mathrm{N}\right\vert_{\boldsymbol{\gamma}=\boldsymbol{\gamma}_0} = 0,
\end{align}
where the derivatives of $K_\mathrm{N}$ are first calculated and then evaluated at the equilibrium point.

\begin{figure}[t!]
	\includegraphics[scale=0.45]{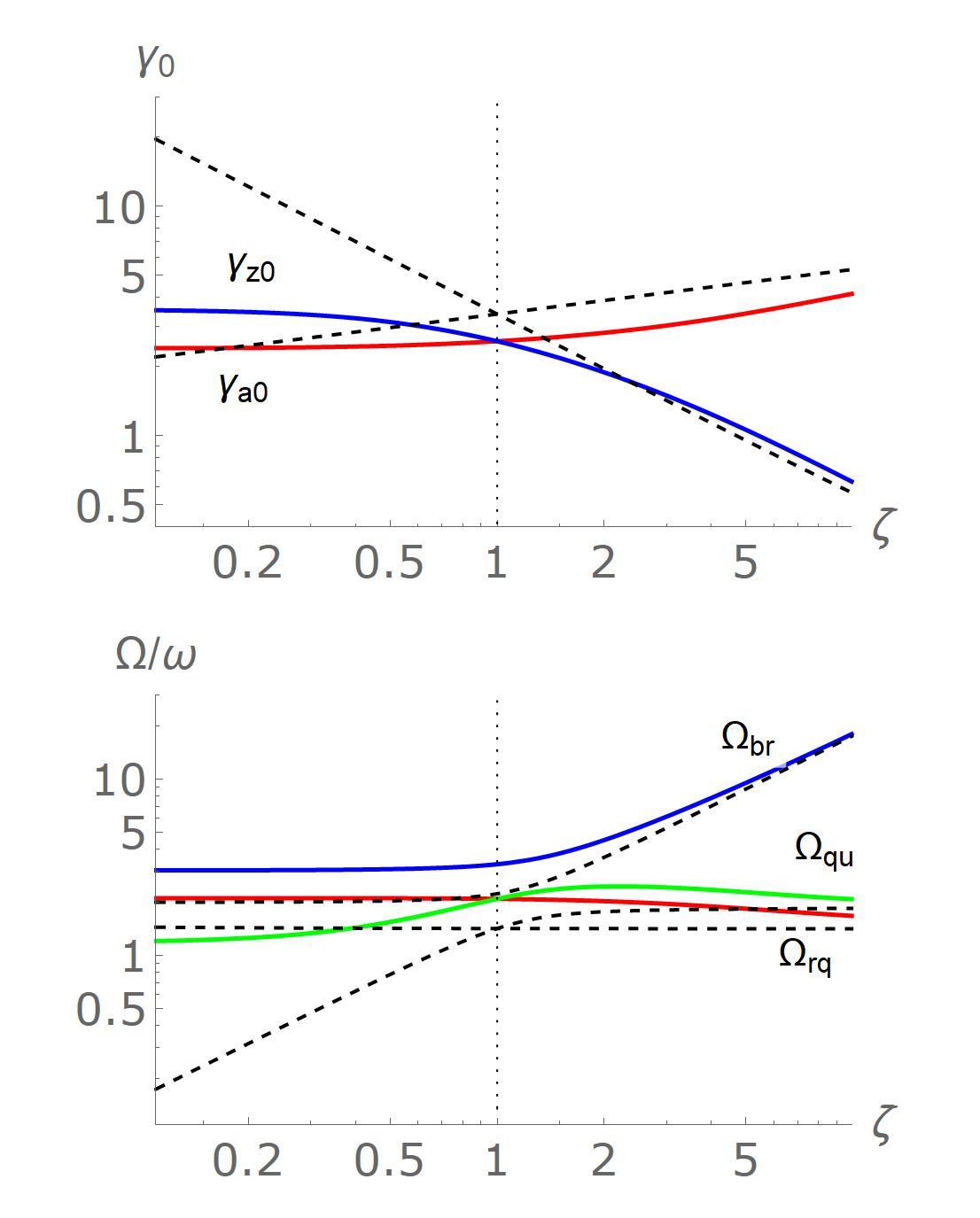}
	\caption{(Colors online) Equilibrium cloud widths $\gamma_{a0}$ and $\gamma_{z0}$, and collective frequencies $\Omega_\mathrm{br}/\omega$, $\Omega_\mathrm{qu}/\omega$, and $\Omega_\mathrm{rq}/\omega$ for an axially symmetric condensate with $P=446$ and $Q=50$ depending on the aspect ratio $\zeta$. The black dashed lines correspond to the case without gravitational interaction, see Fig.~\ref{fig:axial con aspect}.}
	\label{fig:axial Newton aspect}
\end{figure}

Finally, we use the steady state and the function $K_\mathrm{N}$ to derive the contributions to the Hessian matrix due to a Newtonian interaction. These read for the diagonal elements
\begin{align}\label{eq:MN axial diag}
M_\mathrm{N}^{(xx)} &= M_\mathrm{N}^{(yy)} = - \frac{3 Q}{2\sqrt{2\pi}} \nonumber\\
&\times\left.\Bigg[ \frac{3}{4} \gamma_a\;\partial_{\gamma_a}\left(\frac{1}{\gamma_a}\;\partial_{\gamma_a} K_\mathrm{N} \right) + \frac{1}{\gamma_a}\;\partial_{\gamma_a} K_\mathrm{N} \Bigg]\right\vert_{\boldsymbol{\gamma} = \boldsymbol{\gamma}_0}, \\
M_\mathrm{N}^{(zz)} &= - \frac{3 Q}{\sqrt{2\pi}} \nonumber\\
&\times\left.\Bigg[ \gamma_z\;\partial_{\gamma_z}\left( \frac{1}{\gamma_z}\;\partial_{\gamma_z} K_\mathrm{N} \right) + \frac{1}{\gamma_z}\;\partial_{\gamma_z} K_\mathrm{N} \Bigg]\right\vert_{\boldsymbol{\gamma} = \boldsymbol{\gamma}_0}
\end{align}
and for the off-diagonal elements
\begin{align}\label{eq:MN axial off-diag}
M_\mathrm{N}^{(xy)} &= -\frac{3 Q}{8\sqrt{2\pi}}\left.\left[\gamma_a\;\partial_{\gamma_a} \left( \frac{1}{\gamma_a}\;\partial_{\gamma_a} K_\mathrm{N} \right)\right]\right\vert_{\boldsymbol{\gamma} = \boldsymbol{\gamma}_0}, \\
M_\mathrm{N}^{(xz)} &= M_\mathrm{N}^{(yz)} \nonumber\\
&=  -\frac{3 Q}{2\sqrt{2\pi}}\left.\Big[\partial_{\gamma_a}\partial_{\gamma_z} K_\mathrm{N}\Big]\right\vert_{\boldsymbol{\gamma} = \boldsymbol{\gamma}_0}.
\end{align}
Once more, the collective frequencies are then numerically calculated with the eigenvalues of the full Hessian.

\begin{figure}[t!]
	\includegraphics[scale=0.45]{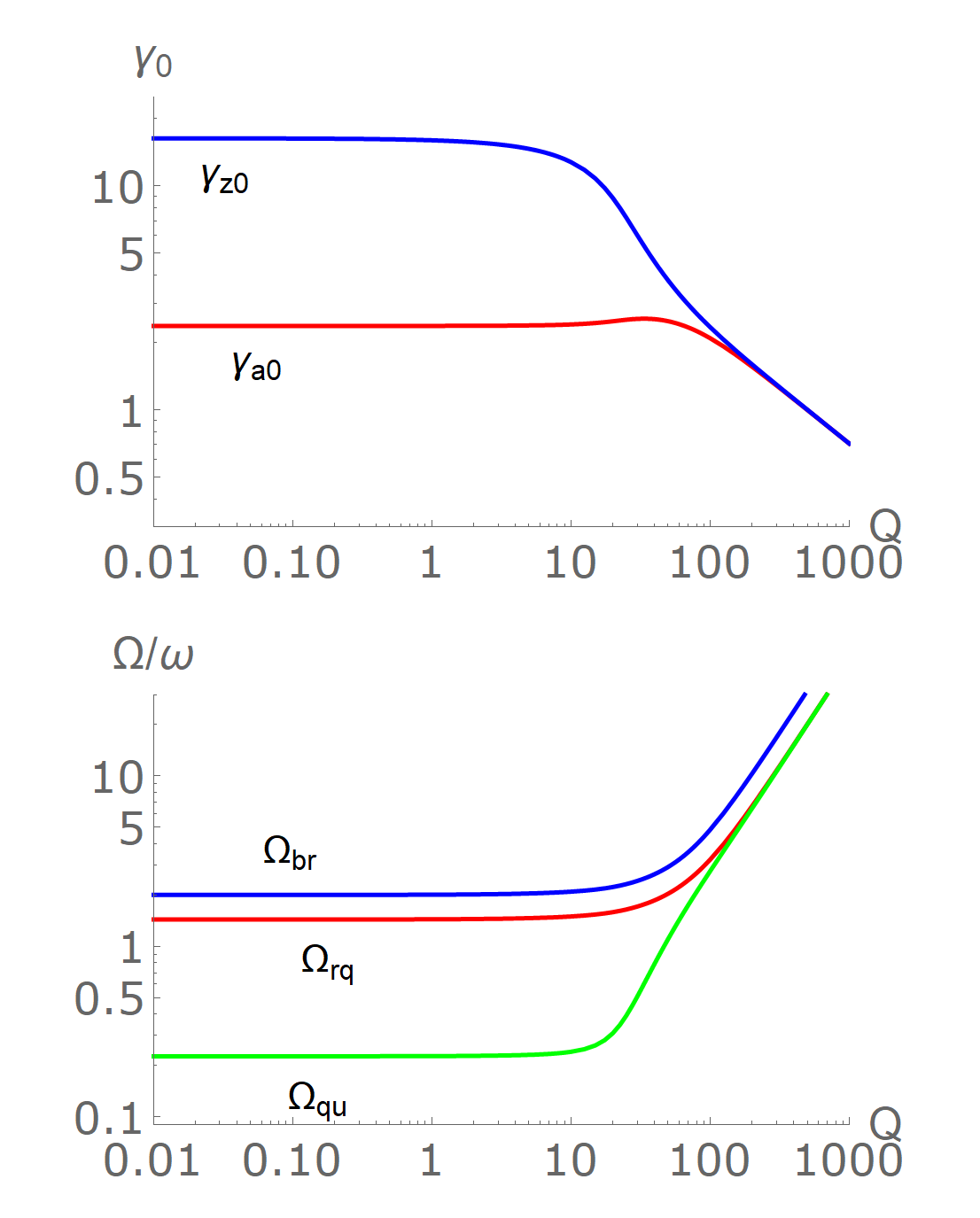}
	\caption{(Colors online) Equilibrium cloud widths $\gamma_{a0}$ and $\gamma_{z0}$, and collective frequencies $\Omega_\mathrm{br}/\omega$, $\Omega_\mathrm{qu}/\omega$, and $\Omega_\mathrm{rq}/\omega$ for a cigar-shaped condensate with $P=446$ and the aspect ratio $\zeta=1/7$ depending on the gravitational interaction strength $Q$.}
	\label{fig:axial Newton cigar}
\end{figure}

Analogous to Sec.~\ref{sec:axial con} we show in Fig.~\ref{fig:axial Newton aspect} the dependency of the equilibrium cloud widths and the collective frequencies on the aspect ratio $\zeta$ for fixed contact interaction strength $P=446$. The black dashed lines indicate the case without gravitational interaction, as seen in Fig.~\ref{fig:axial con aspect}. For realistic and small gravitational interaction strengths the effects are barely visible, so we decided to only show here the plot for $Q=50$. The large gravitational interaction strength now leads in general to smaller equilibrium cloud widths and larger collective frequencies. This is qualitatively similar to the behavior shown in the spherical case in Fig.~\ref{fig:sphere Newton}. Furthermore, the exponential laws for the cloud widths of the case including the contact interaction now break up and the straight lines become curvy. In particular, for disk-shaped condensates the collective frequencies of both the radial quadrupole mode and the out-of-phase quadrupole mode are increased for $\zeta\approx 3$ but again smaller for even larger aspect ratios.

\begin{figure}[t!]
	\includegraphics[scale=0.45]{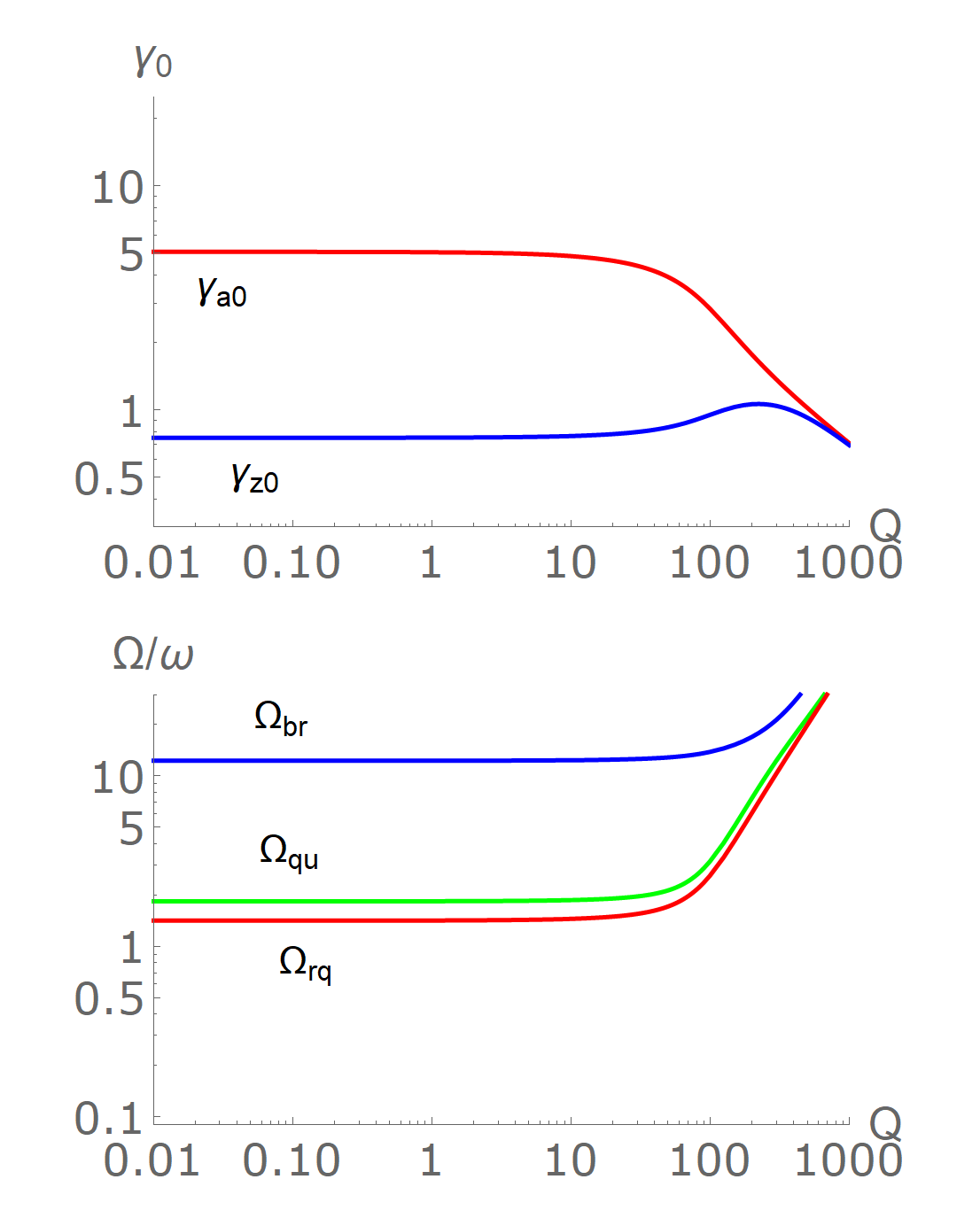}
	\caption{(Colors online) Equilibrium cloud widths $\gamma_{a0}$ and $\gamma_{z0}$, and collective frequencies $\Omega_\mathrm{br}/\omega$, $\Omega_\mathrm{qu}/\omega$, and $\Omega_\mathrm{rq}/\omega$ for a disk-shaped condensate with $P=446$ and the aspect ratio $\zeta=7$ depending on the gravitational interaction strength $Q$.}
	\label{fig:axial Newton disk}
\end{figure}

Next we present the dependency of the cloud widths and the collective frequencies on the gravitational interaction strength for a cigar-shaped condensate with $\zeta=1/7$ in Fig.~\ref{fig:axial Newton cigar} and for a disk-shaped condensate with $\zeta=7$ in Fig.~\ref{fig:axial Newton disk}. Again, for small values of $Q$ the effects of the gravitational interaction are negligible. However, for $Q>10$ gravitational effects become more significant. Interestingly, for very large interactions strength $Q>500$ in both configurations the equilibrium cloud widths coincide, which resembles a spherical form. This is in fact the very nature of Newton's gravitational potential as a conservative potential. If the attractive gravitational force in radial direction significantly surpasses the repulsive contact interaction the condensate transitions to a spherical form. Analogously, the frequencies corresponding to both quadrupole modes also coincide, which shows the degeneracy discussed for the spherical case.
In addition, we notice a global maximum of the lower Gaussian width at $Q\approx 100$ in both cases. The reason for this is the difference in the particle densities in the transversal and longitudinal directions. In case of a cigar-shaped condensate there are on average more particles in the longitudinal direction, which leads to a higher gravitational attraction. This is then compensated due to the repulsive contact interaction by an increase in the size in the transversal direction. This effect is clearly more pronounced in the case of the disc-shaped form, but occurs for larger gravitational strengths than in the cigar-shaped case.

In the end of this section we calculate typical values for the equilibrium widths $\gamma_{a0}$ and $\gamma_{z0}$, and the collective frequencies $\Omega_\mathrm{br}^\mathrm{(N)}$, $\Omega_\mathrm{qu}^\mathrm{(N)}$, and $\Omega_\mathrm{rq}^\mathrm{(N)}$. For a $^{87}\mathrm{Rb}$ condensate we use the interactions strengths $P=446$ and $Q=4\cdot 10^{-19}$, as seen before, with $\omega=1\;\mathrm{kHz}$. For a cigar-shaped condensate with $\zeta=1/7$ we obtain
\begin{align}\label{eq:Newton typ cigar}
\gamma_{a0} &= 1.98\;\mathrm{\mu m},\;\gamma_{z0} = 13.61\;\mathrm{\mu m}, \nonumber\\
\Omega_\mathrm{br}^\mathrm{(N)} &= 2.002\;\mathrm{kHz},\;\Omega_\mathrm{rq}^\mathrm{(N)} = 1.438\;\mathrm{kHz}, \nonumber\\
&\Omega_\mathrm{qu}^\mathrm{(N)} = 0.226\;\mathrm{kHz}
\end{align}
and for a disk-shaped condensate with $\zeta=7$
\begin{align}\label{eq:Newton typ disk}
\gamma_{a0} &= 4.23\;\mathrm{\mu m},\;\gamma_{z0} = 0.63\;\mathrm{\mu m} \nonumber\\
\Omega_\mathrm{br}^\mathrm{(N)} &= 12.290\;\mathrm{kHz},\;\Omega_\mathrm{rq}^\mathrm{(N)} = 1.415\;\mathrm{kHz}, \nonumber\\
&\Omega_\mathrm{qu}^\mathrm{(N)} = 1.839\;\mathrm{kHz}.
\end{align}
Again, these values are a reference for the calculations in the next section.

\subsection{Yukawa interaction}\label{sec:axial Yuk}

In this section we investigate a condensate with a Yukawa-like interaction in cylindrical symmetry. For this, we start with the Lagrangian $L_\mathrm{Yuk}$ given in Eq.~\eqref{eq:LYuk dimless}. After the integration over the polar coordinates $\varphi$ and $\kappa_\rho$ we define analogous to Sec.~\ref{sec:axial Newton} the function
\begin{align}\label{eq:KYuk axial def}
K_\mathrm{Yuk} = \frac{1}{2}&\int_{-\infty}^\infty\mathrm{d}\kappa_z\;\Gamma\left(0,\frac{1}{2}\gamma_a^2 \left(\kappa_z^2+\frac{1}{\bar{\lambda}^2}\right)\right) \nonumber\\
&\times\exp\left\{\frac{1}{2}\gamma_a^2 \left(\kappa_z^2+\frac{1}{\bar{\lambda}^2}\right)\right\}\exp\left\{-\frac{1}{2}\gamma_z^2\kappa_z^2\right\}.
\end{align}
This function now depends on the dimensionless effective range $\bar{\lambda}$. In contrast to $K_\mathrm{N}$ in the Newtonian case the integral in $K_\mathrm{Yuk}$ cannot be solved analytically as far as we know. Nevertheless, we can use the formal definition to derive the two differential equations
\begin{align}\label{eq:ODE Yukawa axial}
\mathrm{i)}\;\; \ddot{\gamma}_a = &-\gamma_a + \frac{1}{\gamma_a^3} + \frac{P}{\gamma_z}\frac{1}{\gamma_a^3} + \frac{3Q}{2\sqrt{2\pi}}\;\partial_{\gamma_a} K_\mathrm{N} \nonumber\\
&+ \frac{3\alpha Q}{2\sqrt{2\pi}}\;\partial_{\gamma_a} K_\mathrm{Yuk}, \\
\mathrm{ii)}\;\; \ddot{\gamma}_z = &-\zeta^2\gamma_z + \frac{1}{\gamma_z^3} + \frac{P}{\gamma_a^2}\frac{1}{\gamma_z^2} + \frac{3Q}{\sqrt{2\pi}}\;\partial_{\gamma_z} K_\mathrm{N} \nonumber\\
&+ \frac{3\alpha Q}{\sqrt{2\pi}}\;\partial_{\gamma_z} K_\mathrm{Yuk},
\end{align}
which then lead to a set of equations determining the equilibrium cloud widths \begin{align}\label{eq:equi width Yukawa axial}
\mathrm{i)}\;\; \gamma_{a0}^5 &- \gamma_{a0} - P\frac{\gamma_{a0}}{\gamma_{z0}} - \gamma_{a0}^4 \frac{3Q}{2\sqrt{2\pi}}\;\left.\partial_{\gamma_a} K_\mathrm{N}\right\vert_{\boldsymbol{\gamma} = \boldsymbol{\gamma}_0} \nonumber\\
& - \gamma_{a0}^4\frac{3\alpha Q}{2\sqrt{2\pi}}\;\left.\partial_{\gamma_a} K_\mathrm{Yuk}\right\vert_{\boldsymbol{\gamma} = \boldsymbol{\gamma}_0} = 0, \\
\mathrm{ii)}\;\; \zeta^2\gamma_{z0}^5 &- \gamma_{z0} - P\frac{\gamma_{z0}^2}{\gamma_{a0}^2} - \gamma_{z0}^4 \frac{3Q}{\sqrt{2\pi}}\;\left.\partial_{\gamma_z}  K_\mathrm{N}\right\vert_{\boldsymbol{\gamma} = \boldsymbol{\gamma}_0} \nonumber\\
& - \gamma_{z0}^4\frac{3\alpha Q}{\sqrt{2\pi}}\;\left.\partial_{\gamma_z}  K_\mathrm{Yuk}\right\vert_{\boldsymbol{\gamma} = \boldsymbol{\gamma}_0} = 0,
\end{align}

\begin{figure}[t!]
	\includegraphics[scale=0.45]{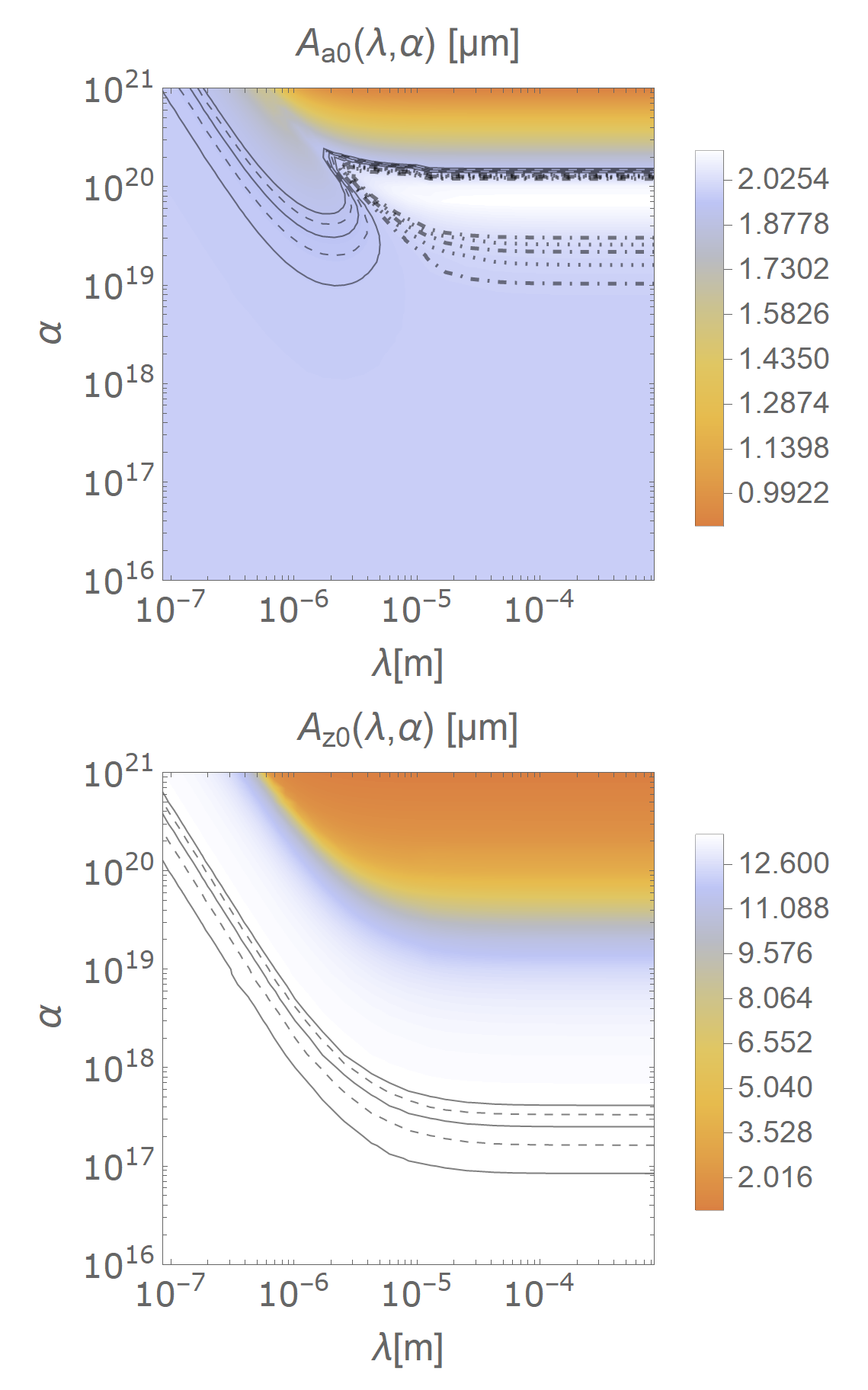}
	\caption{(Colors online) Equilibrium cloud widths $A_{a0}$ and $A_{z0}$ of a cigar-shaped condensate with $P=446$ and $Q=4\cdot10^{-19}$ and the aspect ratio $\zeta=1/7$ depending on the Yukawa parameters $\lambda$ and $\alpha$. The black and black dashed lines indicate a decrease of $0.01\;\mathrm{\mu m}$ to $0.05\;\mathrm{\mu m}$, while the black dotted lines show an increase of $0.01\;\mathrm{\mu m}$ to $0.05\;\mathrm{\mu m}$ compared to the Newtonian case given in Eq.~\eqref{eq:Newton typ cigar}. For better visibility, the curves alternate between solid and dashed lines.}
	\label{fig:cigar Yuk width}
\end{figure}

\begin{figure}[t!]
	\includegraphics[scale=0.45]{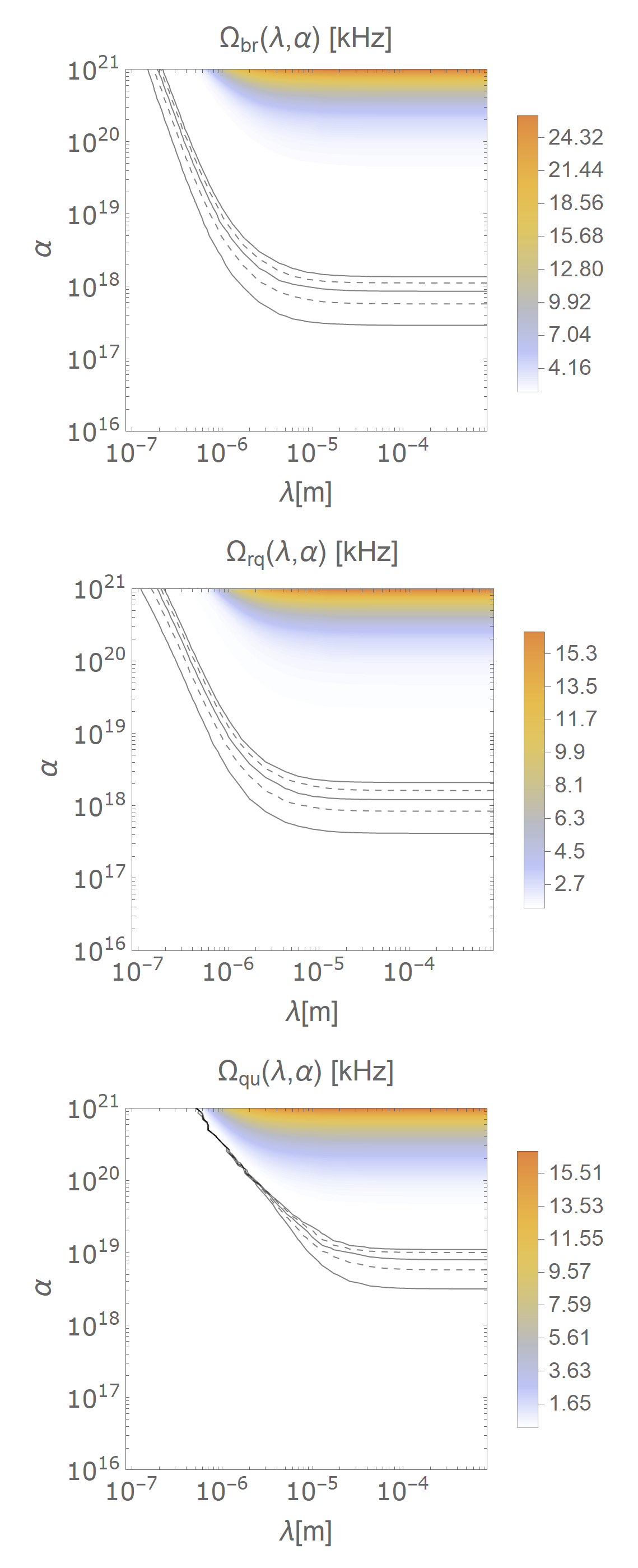}
	\caption{(Colors online) Collective frequencies $\Omega_\mathrm{br}$, $\Omega_\mathrm{qu}$, and $\Omega_\mathrm{rq}$ of a cigar-shaped condensate with $P=446$ and $Q=4\cdot10^{-19}$ and the aspect ratio $\zeta=1/7$ depending on the Yukawa parameters $\lambda$ and $\alpha$. The black and black dashed lines indicate an increase of $1\;\mathrm{Hz}$ to $5\;\mathrm{Hz}$ compared to the Newtonian results given in Eq.~\eqref{eq:Newton typ cigar}. For better visibility, the curves alternate between solid and dashed lines.}
	\label{fig:cigar Yuk frequ}
\end{figure}

Here we have to solve both equations simultaneously with the numerical integration in $K_\mathrm{Yuk}$ to find the equilibrium widths. Once we know the equilibrium, we derive the diagonal elements of the Hessian matrix
\begin{align}\label{eq:MYuk axial diag elements}
&M_{xx} =\;1 + \frac{3}{\gamma_{a0}^4} + \frac{2P}{\gamma_{a0}^4\gamma_{z0}} \nonumber\\
&- \frac{3Q}{2\sqrt{2\pi}}\left.\Bigg[ \frac{3}{4} \gamma_a\;\partial_{\gamma_a}\left(\frac{1}{\gamma_a}\;\partial_{\gamma_a} K_\mathrm{N} \right) + \frac{1}{\gamma_a}\;\partial_{\gamma_a} K_\mathrm{N} \Bigg]\right\vert_{\boldsymbol{\gamma} = \boldsymbol{\gamma}_0} \nonumber\\
&- \frac{3\alpha Q}{2\sqrt{2\pi}}\left.\Bigg[ \frac{3}{4} \gamma_a\;\partial_{\gamma_a}\left(\frac{1}{\gamma_a}\;\partial_{\gamma_a} K_\mathrm{Yuk} \right) + \frac{1}{\gamma_a}\;\partial_{\gamma_a} K_\mathrm{Yuk} \Bigg]\right\vert_{\boldsymbol{\gamma} = \boldsymbol{\gamma}_0}, \\
&M_{zz} = \zeta^2 + \frac{3}{\gamma_{z0}^4} + \frac{2P}{\gamma_{a0}^2\gamma_{z0}^3} \nonumber\\
&- \frac{3Q}{\sqrt{2\pi}}\left.\Bigg[ \gamma_z\;\partial_{\gamma_z}\left( \frac{1}{\gamma_z}\;\partial_{\gamma_z} K_\mathrm{N} \right) + \frac{1}{\gamma_z}\;\partial_{\gamma_z} K_\mathrm{N} \Bigg]\right\vert_{\boldsymbol{\gamma} = \boldsymbol{\gamma}_0} \nonumber\\
&-\frac{3\alpha Q}{\sqrt{2\pi}}\left.\Bigg[ \gamma_z\;\partial_{\gamma_z}\left( \frac{1}{\gamma_z}\;\partial_{\gamma_z} K_\mathrm{Yuk} \right) + \frac{1}{\gamma_z}\;\partial_{\gamma_z} K_\mathrm{Yuk} \Bigg]\right\vert_{\boldsymbol{\gamma} = \boldsymbol{\gamma}_0}
\end{align}
as well as the off-diagonal elements
\begin{align}\label{eq:MYuk axial off-diag elements}
M_{xy} &= \frac{P}{\gamma_{a0}^4\gamma_{z0}} - \frac{3Q}{8\sqrt{2\pi}}\left.\left[\gamma_a\;\partial_{\gamma_a} \left( \frac{1}{\gamma_a}\;\partial_{\gamma_a} K_\mathrm{N} \right)\right]\right\vert_{\boldsymbol{\gamma} = \boldsymbol{\gamma}_0} \nonumber\\
&-\frac{3\alpha Q}{8\sqrt{2\pi}} \left.\left[\gamma_a\;\partial_{\gamma_a} \left( \frac{1}{\gamma_a}\;\partial_{\gamma_a} K_\mathrm{Yuk} \right)\right]\right\vert_{\boldsymbol{\gamma} = \boldsymbol{\gamma}_0}, \\
M_{xz} &= \frac{P}{\gamma_{a0}^3\gamma_{z0}^2} - \frac{3Q}{2\sqrt{2\pi}}\;\left.\Big[\partial_{\gamma_a}\partial_{\gamma_z} K_\mathrm{N}\Big]\right\vert_{\boldsymbol{\gamma} = \boldsymbol{\gamma}_0} \nonumber\\
&-\frac{3\alpha Q}{2\sqrt{2\pi}}\left.\Big[\partial_{\gamma_a}\partial_{\gamma_z} K_\mathrm{Yuk}\Big]\right\vert_{\boldsymbol{\gamma} = \boldsymbol{\gamma}_0}
\end{align}
for a cylindrically symmetric condensate, where the particles interact via a contact, Newtonian, and a Yukawa-like potential with each other.

\begin{figure}[t!]
	\includegraphics[scale=0.45]{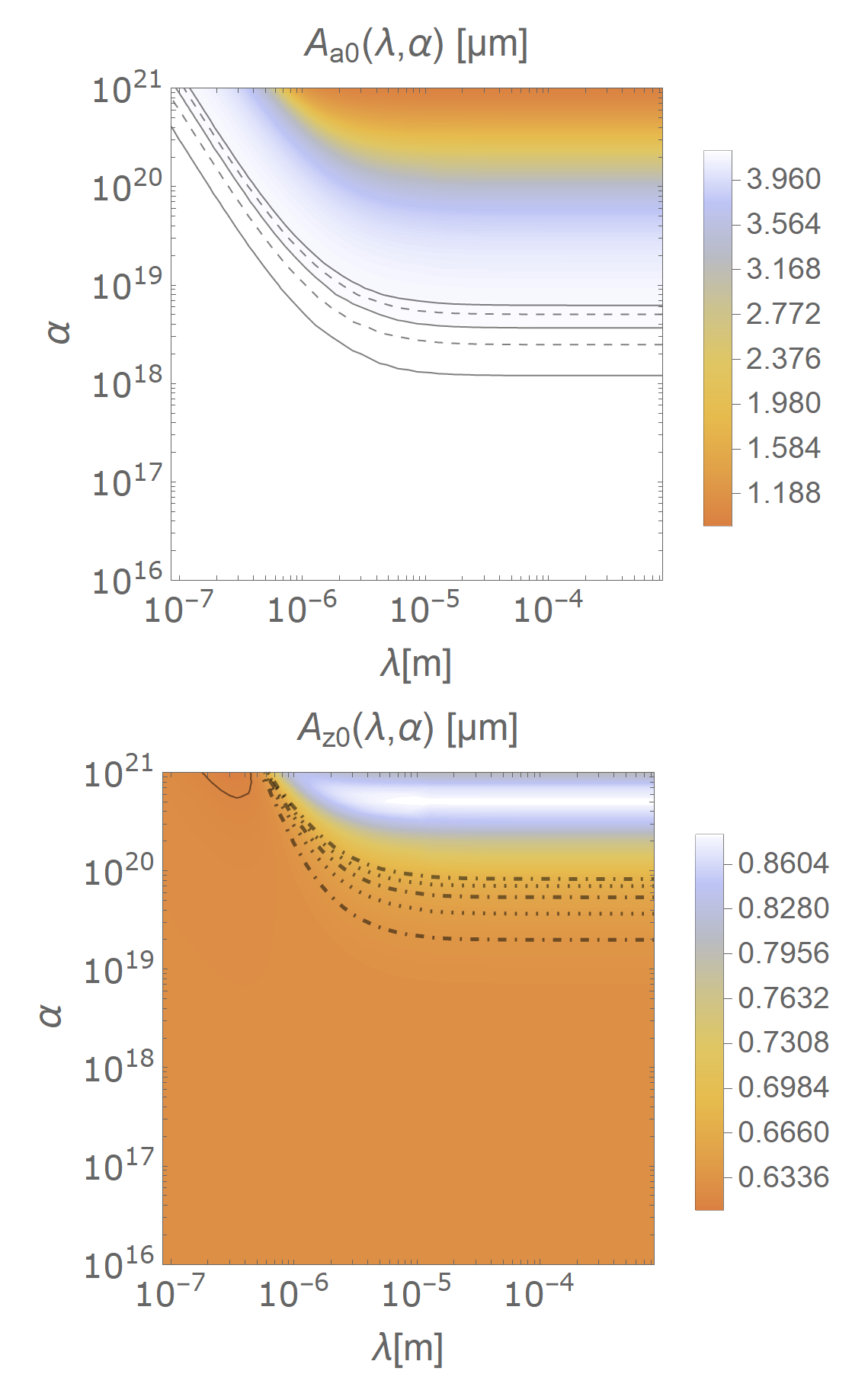}
	\caption{(Colors online) Equilibrium cloud widths $A_{a0}$ and $A_{z0}$ of a disk-shaped condensate with $P=446$ and $Q=4\cdot10^{-19}$ and the aspect ratio $\zeta=7$ depending on the Yukawa parameters $\lambda$ and $\alpha$. The black and black dashed lines indicate a decrease of $0.01\;\mathrm{\mu m}$ to $0.05\;\mathrm{\mu m}$, while the black dotted lines show an increase of $0.01\;\mathrm{\mu m}$ to $0.05\;\mathrm{\mu m}$ compared to the Newtonian case given in Eq.~\eqref{eq:Newton typ disk}. For better visibility, the curves alternate between solid and dashed lines.}
	\label{fig:disk Yuk width}
\end{figure}

\begin{figure}[t!]
	\includegraphics[scale=0.45]{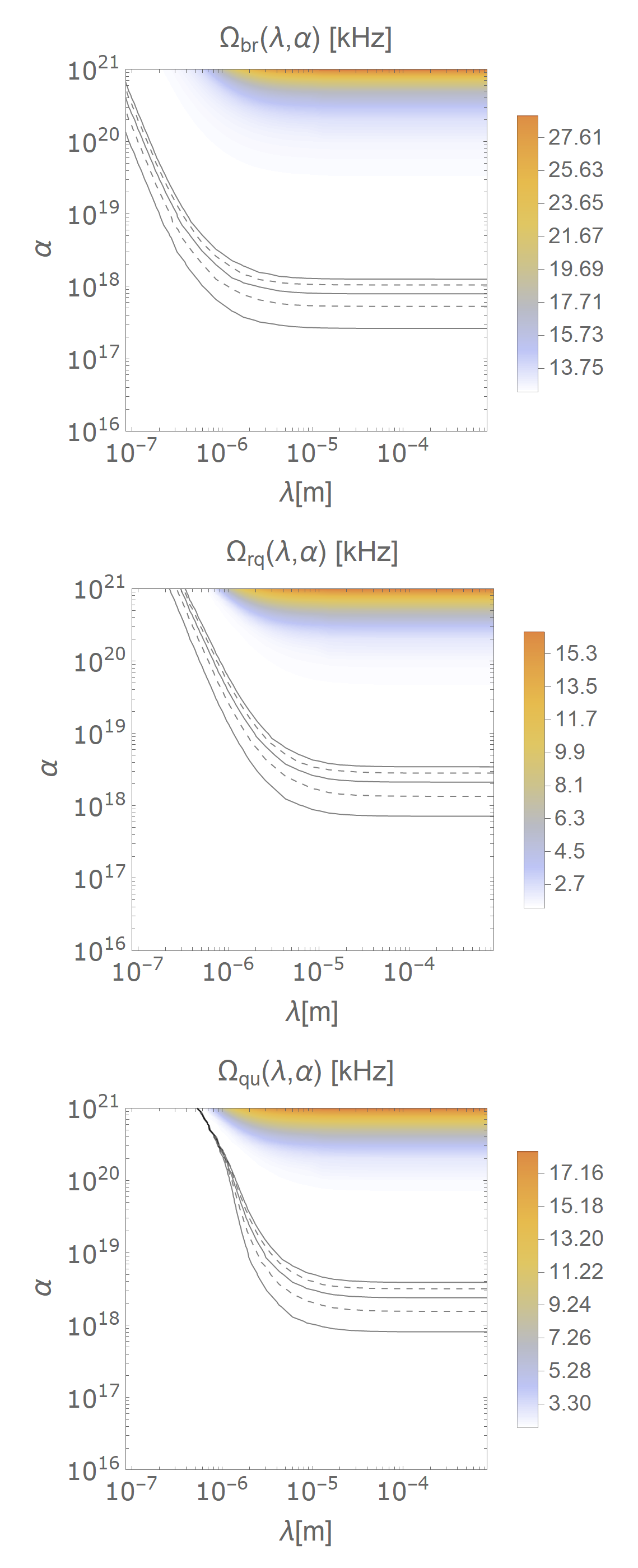}
	\caption{(Colors online) Collective frequencies $\Omega_\mathrm{br}$, $\Omega_\mathrm{qu}$, and $\Omega_\mathrm{rq}$ of a disk-shaped condensate with $P=446$ and $Q=4\cdot10^{-19}$ and the aspect ratio $\zeta=7$ depending on the Yukawa parameters $\lambda$ and $\alpha$. The black and black dashed lines indicate an increase of $1\;\mathrm{Hz}$ to $5\;\mathrm{Hz}$ compared to the Newtonian results given in Eq.~\eqref{eq:Newton typ disk}. For better visibility, the curves alternate between solid and dashed lines.}
	\label{fig:disk Yuk frequ}
\end{figure}

In Fig.~\ref{fig:cigar Yuk width} we show the results for the equilibrium cloud widths for a cigar-shaped condensate with the aspect ratio $\zeta=1/7$. Like before we set $P=446$ and $Q=4\cdot10^{-19}$. The values for small interactions strengths $\alpha$ are similar to those derived for the contact interaction, which we discussed in Sec.~\ref{sec:axial con}, see also Fig.~\ref{fig:axial con aspect} at the specific value of the aspect ratio. For increased interaction strength both widths are overall decreased, visible by the color code of the contour plot. In case of the transversal Gaussian width $A_{a0}$ there is also an area visible, where the width is first slightly increased and then again decreased for higher values of $\alpha$. Note that for a cigar-shaped condensate the transversal width is always smaller than the longitudinal width, such that this resembles our results for the Newtonian interaction in Fig.~\ref{fig:axial Newton cigar}.
The corresponding collective frequencies to the equilibrium cloud widths are shown in Fig.~\ref{fig:cigar Yuk frequ}. At first sight the results look quite similar to the spherical case presented in Fig.~\ref{fig:sphere Yuk frequ}. Notably, the frequencies of both quadrupole modes now differ as expected for cylindrical symmetry. A comparison between the cigar-shaped and the spherical case will be shown later also including the disk-shaped form.

For a disk-shaped condensate with $\zeta=7$ the results for the equilibrium widths are shown in Fig.~\ref{fig:disk Yuk width} and for the collective frequencies in Fig.~\ref{fig:disk Yuk frequ}. Again, we see similar results, however, the equilibrium cloud widths are interchanged, since $A_{z0}$ is now the smaller width.

\begin{figure}[t!]
	\includegraphics[scale=0.45]{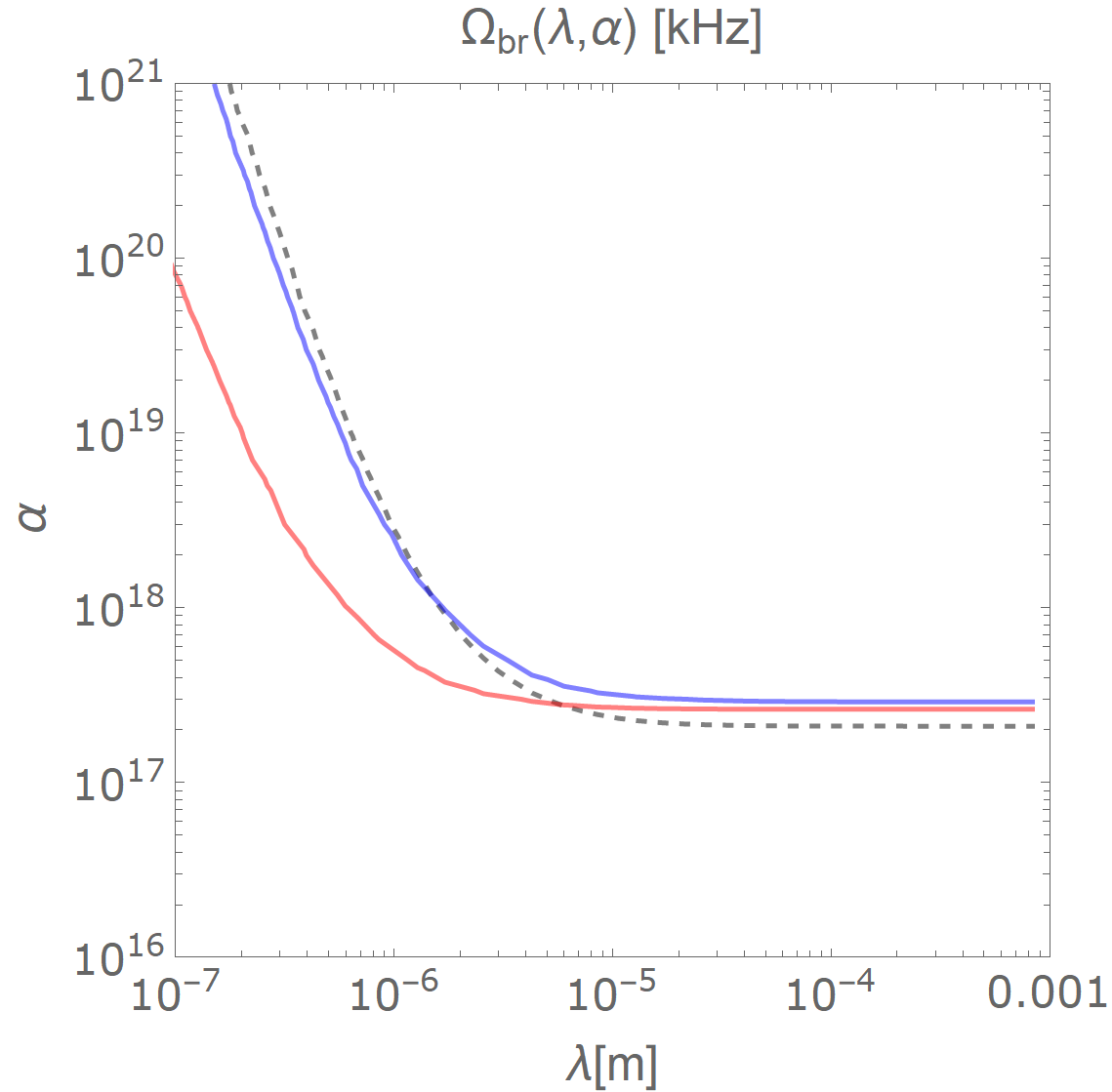}
	\caption{(Colors online) Comparison of the collective frequencies $\Omega_\mathrm{br}$ of the breathing mode between a cigar-shaped condensate with $\zeta=1/7$ (blue), a disk-shaped condensate with $\zeta=7$ (red) and a spherical condensate (black dashed) as a function of the Yukawa parameters $\lambda$ and $\alpha$. The lines show the constraints for a frequency difference of $1\;\mathrm{Hz}$ compared to the corresponding Newtonian case given in Eq.~\eqref{eq:Newton typ cigar} and Eq.~\eqref{eq:Newton typ disk}.}
	\label{fig:comp Yuk sphere axial}
\end{figure}

Finally, we compare our results of the cigar-shaped and disk-shaped condensates with the spherical case discussed previously. Since the collective frequency of the breathing mode leads to the best constraints for each case, we restrict the comparison to this frequency. The frequency of the breathing mode in all three configurations are shown in Fig.~\ref{fig:comp Yuk sphere axial}. For a clear picture we only include the constraints for an accuracy of $1\;\mathrm{Hz}$ of the difference to the Newtonian case. In the figure we see that the results of the spherical and cigar-shaped case with the aspect ratio $\zeta=1/7$ are very close to each other. In particular, for the cigar-shaped condensate the constraints for the effective range $\lambda$ are slightly better, although we lose a little bit of the precision of the interaction strength $\alpha$. However, in case of a disk-shaped condensate with $\zeta=7$, we see a significant improvement for the constraints of the effective range. In regard to the experimental data given in Ref.~\cite{Murata}, also shown in Fig.~\ref{fig:sphere Yuk frequ comp}, it is clear that improving the constraints for $\lambda$ is more important than for $\alpha$. So as a final result, the quasi effective two dimensional disk-shaped condensates seems better suited in that regard.

\begin{figure}[t!]
	\includegraphics[scale=0.45]{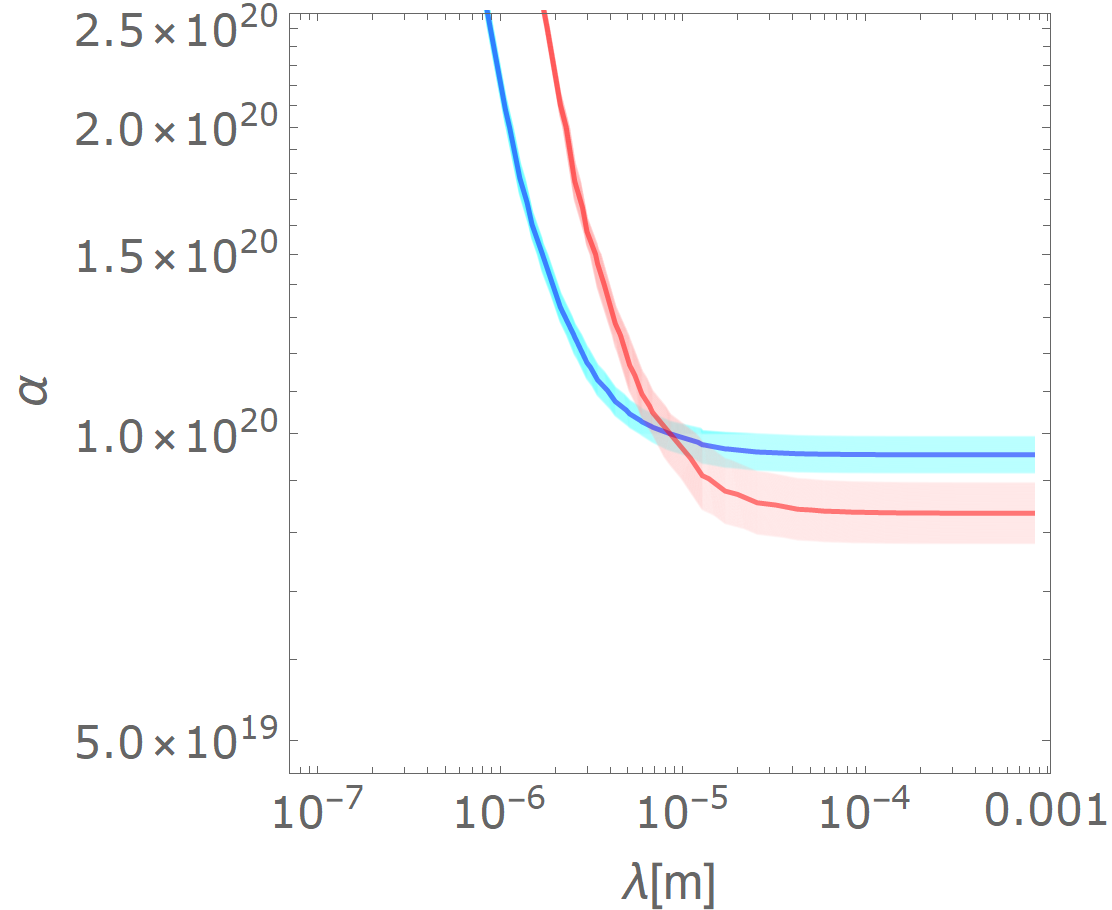}
	\caption{Determination of $\alpha$ and $\lambda$ with the collective frequencies of the breathing mode (blue) and the our-of-phase quadrupole mode (red). The light blue and red shaded area denotes an error in the frequency of $20\;\mathrm{Hz}$. The intersection area of both curves lead to the values $\alpha$ and $\lambda$ including error bars.}
	\label{fig:det alpha lambda}
\end{figure}

In the last part, we propose a method for theoretically reconstruction the Yukawa parameters based on experimental measurements of at least two different collective frequencies. For this purpose, we consider a disk-shaped $^{87}\mathrm{Rb}$ condensate with the aspect ratio $\zeta=7$, a fixed particle number $N=10^5$, and the frequency scale $\omega=1\;\mathrm{kHz}$. Thus the interactions strengths are $P=446$ and $Q=4\cdot 10^{-19}$, as we have seen previously. Now let us assume for a moment that the Yukawa parameters are known, e.g. $\alpha=10^{20}$ and $\bar{\lambda}=10$, which corresponds in our example to $\lambda=8.53\;\mathrm{\mu m}$. Using the Eqs.~\eqref{eq:equi width Yukawa axial} we can calculate the equilibrium cloud widths and use the elements of the Hessian matrix in Eqs.~\eqref{eq:MYuk axial diag elements} and~\eqref{eq:MYuk axial off-diag elements} to derive the collective frequencies of the breathing mode and the quadrupole modes. We now discuss only one quadrupole mode, namely the out-of-phase quadrupole mode, since both are quite similar and overlap. In our example, we obtain $\Omega_\mathrm{br}=12.752\;\mathrm{kHz}$ and $\Omega_\mathrm{qu} = 2.023\;\mathrm{kHz}$. Now, pretending that we have measured these values, we look for the corresponding contour lines for these frequencies, i.e., in Fig.~\ref{fig:disk Yuk frequ}. Both curves are shown in Fig.~\ref{fig:det alpha lambda}. At the intersection point, we obtain independently determined values for $\alpha$ and $\lambda$. Of course, in our example, we correctly recover the values for the Yukawa parameters that we initially assumed. Furthermore, we include in Fig.~\ref{fig:det alpha lambda} an hypothetical error of $20\;\mathrm{Hz}$ for the frequencies. Thus we estimate the errors $\Delta\lambda\approx 3\;\mathrm{\mu m}$ and $\Delta\alpha\approx  0.1\cdot 10^{20}$. Obviously, this is just an example and serves as a proof of principle, since the chosen values for $\alpha$ and $\lambda$ are already excluded by experimental data. Nevertheless, the idea is to measure the collective frequencies with such accuracy that the error bars do not overlap for large effective ranges. As shown in our example, one could then determine the Yukawa parameters by the intersection point of at least two contour lines.

\section{Conclusions}\label{sec:summary}

In regard to the question of non-Newtonian gravity at short ranges, we propose in this work a hypothetical model of a self-gravitating Bose-Einstein condensate as an additional test. For this, we assume that the particles within the condensate interact via a Newtonian or a Yukawa-like potential as the most common parametrization of models of modified gravity. Using a variational ansatz, we analytically derive corrections to the collective frequencies in a spherical condensate caused by the two gravitational interactions. In case of the Newtonian potential these corrections are twenty orders of magnitude smaller than the contact interaction and therefore, as expected, insignificant for realistic condensates. However, assuming a fifth force, i.e., in form of a Yukawa potential this gap in the interaction strength is compensated by a Yukawa parameter. As a consequence, even for a common condensate of $^{87}\mathrm{Rb}$ atoms we calculate notable corrections to the collective frequencies compared to those in the Newtonian case. Based on these correction, we calculate constraints for both Yukawa parameters the interaction strength and the effective range. It is shown that increasing the mass of the particles, the particle number, and the trapping frequency as well as decreasing the s-wave scattering length, leads to an improvement of the constraints. For a realizable $^{174}\mathrm{Yb}$ condensate, our results are close to experimentally verified data from other tests and even slightly improves the constraints, as shown in Fig.~\ref{fig:sphere Yuk frequ comp}. Additionally, we consider axially symmetric condensates to further improve our results. In fact, for a disk-shaped condensate the constraints derived from the collective frequency of the breathing mode, in particular the constraints for the effective range, are significantly better than in the spherical case. We also show by an example that we can in principle determine the Yukawa parameters independently by a measurement of at least two collective frequencies.

So far, this is a toy model based on an analytical approximate solution of the Gross-Pitaevskii equation. A numerical and experimental verification would be interesting as well as a rigorous derivation from first principles. In addition, other modes such as the two-dimensional scissor modes could be investigates, since the disk-shaped case has provided the best constraints so far. In terms of gravity other models of modified gravity could also be considered with a prominent example being chameleon fields. There, one assumes an additional scalar field, where the gravitational coupling depends directly on the matter density. In fact, the effects of such a chameleon field are larger for smaller densities. Thus, for dilute Bose-Einstein condensates one expects larger effects. However, given many experimental possibilities in the field of quantum gases as well as countless theoretical modifications in the area of short-range gravity, we hope that this work offers a playground for improvements in future projects.

We would like to thank A. Pelster and E. Stein for useful discussion. This work was funded by the Deutsche Forschungsgemeinschaft (DFG, German Research Foundation) with CRC 1227 DQ-mat -- 274200144 -- with project B08 and under Germany's Excellence Strategy -- EXC-2123 "QuantumFrontiers" -- 390837967.

\end{document}